\documentclass[prd,aps,floats,floatfix,eqsecnum,nofootinbib,12pt]{revtex4-1}
\usepackage{verbatim,graphicx,amssymb,amsbsy,bm,amsmath,rotating,epsfig}
\usepackage{psfrag}
\usepackage{hyperref}
\usepackage{fontenc}
\usepackage[latin1]{inputenc}
\usepackage{pstricks,pst-node,pst-text,pst-3d}
\usepackage{textcomp}

\newcommand{\be}{\begin{equation}}
\newcommand{\ee}{\end{equation}}
\newcommand{\bea}{\begin{eqnarray}}
\newcommand{\eea}{\end{eqnarray}}

\begin{document}
\title{Coherent States of Quantum Space-Times \\
for Black Holes and de Sitter space-time}
\author{Diego J. Cirilo-Lombardo $^{(a)}$}
\author{Norma G. Sanchez $^{(b)}$}
\affiliation{${(a)}$ M. V. Keldysh Institute of the Russian Academy of Sciences, Federal Research Center-Institute of Applied Mathematics, Miusskaya sq. 4, 125047 Moscow, Russian Federation and CONICET-Universidad de Buenos Aires, Departamento de Fisica, Instituto de Fisica Interdisciplinaria y Aplicada (INFINA), Buenos Aires, Argentina. 
\\
${(b)}$ International School of Astrophysics Daniel Chalonge - Hector de Vega, CNRS, INSU-Institut National des Sciences de l'Univers, Sorbonne University, 75014 Paris, France.}
\date{\today }

\begin{abstract}
\textbf{Abstract:} We provide a group theory approach to coherent states describing quantum space-time and its properties. This provides a relativistic framework for the metric of a Riemmanian space with bosonic and fermionic coordinates, its  continuum and discrete states, and a kind of {\it"quantum optics"} for the space-time. {\bf New} results of this paper are: (i)  The space-time is described as a physical coherent state of the complete covering of the SL(2C) group, eg the Metaplectic group Mp(n). (ii) (The discrete structure arises from its two irreducible: \textit{even} $(2n)$  and \textit{odd} $(2n\;+\;1)\;$   
representations,  ($n = 1,\, 2, \,3\,...$ ), spanning  the complete Hilbert space $\mathcal{H} = \mathcal{H}
_{odd}\oplus \mathcal{H}_{even}$. Such a global or {\it complete} covering guarantees the CPT symmetry and unitarity. Large $n$ yields
the classical and continuum manifold, as it must be. (iii) The coherent and  squeezed states and Wigner functions of quantum-space-time for black holes and de Sitter, and (iv) for the quantum space-imaginary time (instantons), black holes in particular.  They encompass the semiclassical space-time behaviour plus high quantum phase oscillations, and notably account for the classical- quantum gravity duality and trans-Planckian domain. The Planck scale consistently corresponds to the coherent state eigenvalue $\alpha = 0$ (and to the $n = 0$ level in the discrete representation).  It is remarkable the power of coherent states in describing both continuum and discrete space-time.  The quantum space-time description is {\it regular}, there is no any space-time singularity here, as it must be. 
\newline

(b): Norma.Sanchez@orange.fr \\
 \url{https://chalonge-devega.fr/sanchez/}

\end{abstract}


\maketitle

\tableofcontents

\section{Introduction and Results} \label{introduction}

Quantum space-time is a key concept both for quantum theory in its own and for a
quantum gravity theory. Coherent states are a fundamental part in quantum physics with
multiple theoretical and practical realizations from mathematical physics to
quantum optics and wave packet experiments, see for example Refs \cite{Kla1} to \cite{Kla2}
and refs therein. In this paper, within a group theory approach, we construct generalized coherent states to describe quantum space-time.

\medskip

We describe quantum space-time as arising from a mapping
$P(G,\mathcal{M})$ between the quantum phase space manifold of a group $\mathcal{G}$ and the real space-time
manifold $\mathcal{M}$. The metric $g_{ab}$ on the phase space group manifold determines the space-time
metric of $\mathcal{M}$ after identification of one component of the momentum $P$ operator with the time $T$.
The signature of the metric depends on the compact or non compact nature of the group,
but in the most cases of physical interest, the real space-time signature and its hyperbolic structure require non compact groups.

\medskip

A  group theory approach, a quantum algebra, reveals a key part in the quantum space-time description in order to obtain the line element associated to a discrete quantum structure of the space-time. Such an emergent metric is obtained here from a Riemmanian phase space and described as a physical coherent state of
the underlying covering of the group SL(2C): Interestingly, it  appears
necessary to consider the {\it complete covering} of the symplectic group, that is the
Metaplectic group $Mp(n)$, its spectrum for all $n$ leading in particular for very large $n$ the continuum space-time. 
This approach allows us to construct here coherent states of the coset type for the quantum space-time and describe with them coherent de Sitter and 
 black hole states. 

\medskip
 
This quantum description is
based on the phase space of a relativistic particle in the superspace with
bosonic and fermionic coordinates, allowing to conserve at the quantum level
the square root forms of the geometrical operators (eg the Hamiltonian or
Lagrangian). 
The discrete spacetime structure arises from the basic states
of the Metaplectic representation with one interesting feature to remark
here: The decomposition of the SO(2,1) group into two irreducible
representations span  \textit{even} $\left\vert \;2n\;\right\rangle $ and \textit{odd} $\left\vert \;2n \;+\;1\;\right\rangle$ states, ($n = 1,\, 2, \,3\,...$ ),
respectively, whose totality is covered by the  Metapletic group. In the Metaplectic
representation the general or {\it complete} states must be the sum of the \textit{two} kind of states : even
and odd $n$ states spanning respectively the two Hilbert sectors $\mathcal{H}%
_{1/4}$ and $\mathcal{H}_{3/4}$, whose complete covering is $\mathcal{H}%
_{1/4}\oplus \mathcal{H}_{3/4}$. This yields the relativistic quantum
space-time metric with discrete structure . For increasing number of levels $n$,
the metric solution goes to the continuum and to a classical manifold as it must be. Such a  global or
{\it complete} covering with the sum of the two sectors, even and odd states to have the complete Hilbert space reflects the CPT symmetry and unitarity of the description. 

As we know, the Metaplectic group $ M_p(2)$ acts irreducibly on each of the subspaces $%
\mathcal{H}_{1/4},$ $\mathcal{H}_{3/4}$ (even and odd) by which the total
Hilbert space (namely $\mathcal{H}$ ) is divided according to the Casimir
operator: 
\begin{equation*}
K^{2} \;= \; K_{3}^{2}\;-\;K_{1}^{2}\;-\;K_{2}^{2}\;= \;k \;\left(\, k\,-\,1\,\right) \; = \;- \, \frac{3}{16}\;\mathbb{%
I}
\end{equation*}
giving precisely the values $k \; = \;1/4,\;\, 3/4$. Then,
\be
\mathcal{H}_{1/4}  \;= \; \text{Span}\left\{ \; \left\vert \;n\; \text{even}\;\right\rangle
: \;n\; = \;0,\,2,\,4,\,6,\,...\;\right\}
\ee
\be
\mathcal{H}_{3/4}\; = \; \text{Span} \left\{ \;\left\vert \;n\; \text{odd}\;\right\rangle
:\;n\;= \;1,\,3,\,5,\,7,\,....\;\right\}
\ee

Based on the highest eigenvalue of  
 the number operator $T_{3}\,\left\vert \,
n\,\right\rangle \, = \,-\,\frac{1}{2}\left( \,n\,+ \,\frac{1}{2}\,\right) \left\vert
\,n\,\right\rangle $ occurring in $\mathcal{H} \;\equiv \;\mathcal{H}_{1/4}\oplus 
\mathcal{H}_{3/4}$, the two unitary irreducible representations (UIR) $\mathcal{D}$ of $%
Mp(2)$ are denoted as: \begin{equation}
(UIR) \quad \text{ restricted \;to \;}\mathcal{H}_{1/4}  \quad \rightarrow  \quad \mathcal{D}_{1/4} \; \in \; \;
Mp(2) \\
\end{equation}
\begin{equation}
(UIR)\quad \text{ restricted \;to \;}\;\mathcal{H}_{3/4} \;\rightarrow \quad  \mathcal{D}_{3/4} \quad \in \;
Mp(2)
\end{equation}

\medskip

One of the  clear examples of the group theory approach presented here is the
quantum space-time derived from the phase space of the harmonic oscillator  (Refs  \cite{Sanchez2}, \cite{NSPRD2021}, \cite {NSPRD2023}), and the mapping $(X, P ) \rightarrow (X, T )$, in the case of the inverted (imaginary frequency) oscillator, or alternatively $\rightarrow (X, iT )$,  in the  normal (real frequency) oscillator. The inverted oscillator in its different representations does appear in a variety
of interesting physical situations from particle physics to black holes and modern cosmology as inflation and today dark energy, eg Refs \cite{dVSanchez} to \cite{deVegaSanchez2007}. 

The group theory framework  presented here to describe  quantum space-time and its coherent states allows to correlate and extend the approachs of Refs \cite{Sanchez2}, \cite{NSPRD2021}, \cite {NSPRD2023} and  \cite{DiegoJMP}, \cite {DiegoHO}, \cite {DiegoHO1}  to obtain {\bf new} results. {\bf Novel results} of this paper are:

{\bf(i)} the generalization of the quantum light-cone to include  fermionic coordinates, 

{\bf(ii)} the construction of coherent and squeezed states of quantum space-time, their properties and interpretation, their continuum and discrete representations, and  for both, de Sitter and black hole space-times.  

{\bf(iii)} The coherent states for the quantum space-imaginary time instantons, for  black holes in particular.

{\bf(iv)} We find that coherent states encompass the space-time behaviour in the 
semiclassical and classical de Sitter and  black hole regions, exhibit high quantum phase oscillations of the space-time, and  account for the classical-quantum gravity duality  and the trans-Planckian scales.

\begin{itemize}
\item{ It is remarkable the power of coherent states in describing both continuum and discrete space-time, {\it even} in the Planckian and trans-Planckian domains:}

\item{The Planck scale consistently corresponds here to the continuum coherent state eigenvalue $\alpha = 0$, (and to the fundamental state $n = 0$ in the discrete representation). Higher values of $\alpha $ in the quantum gravity (trans-Planckian) domain account for the smaller and sub-Planckian sizes and higher excitations.} 

\item{One of the  {\bf new} features, for the space-imaginary time instantons is the emergence of a {\it maximum eigenvalue} $\alpha$ 
characterizing the coherent states due to the minimal non-zero quantum radius  because of the  minimal quantum uncertainty $\Delta\,X \,\Delta\,T = \hslash\,/\,2$, in particular in the central and {\it regular}  black hole quantum region. The coherent state instanton remarkably accounts for this quantum gravity feature and determines the radius being \;$$\mathcal{R}_{0} \,(\,l_P,\, t_P \,)^{\,2}\;\;=\;\;
\frac{1}{\sqrt{\,\pi}}\;\left[\;\frac{1}{l_P } \;+\; 
\frac{l_P}{\hslash}\;\right], $$\; 
$l_P$ being the Planck length. The origin is flurried or smoothed within this constant  and bounded curvature region.} 

\item{In the quantum space-time description, there is no any space -time singularity as it must be. The consistent description by coherent states of such quantum scales does appear here as a result of the classical-quantum gravity duality across the Planck scale, and reflected  here in the double covering of the SL(2C) group or Metaplectic  symmetry.}
\end{itemize}

\textbf{This paper is organized as follows:} In Section (II) we construct the 
generalized or coset group coherent states and squeezed states and the corresponding Wigner quasiprobality functions.
In Section (III), we describe the Mp(n) general group approach including bosonic and fermionic coordinates, in particular Mp(2) and the geometrical interpretation of high quantum oscillatory effects in this context.  Section (IV) describes the Mp (n) associated relativistic wave equation, the complete Hilbert space and the discrete representations. The physical states, the Mp(2) squeezed vacuum  and the direct sum of the both odd and even states, necessary to uncover the complete space-time are discussed in this section. In Section (V) we find the coherent states for the de Sitter and  black hole space-times, their properties 
and interpretation. Section (VI) deals with the coherent states of quantum (imaginary time) instantons, for black holes in particular and its new effects. 
In Section (VII) provides a discussion in the context of our results and other Refs, and Section  VIII summarizes our remarks and conclusions.

\section{Quantum Coherent States} \label{coherentstates}

We construct first coherent states within a group theory approach of the
Klauder-Perelomov type (Refs \cite{Kla},  \cite{pere}) or coset group coherent states and then in Sections  and  we describe them in terms of the Metaplectic Mp(n) group and the associated relativistic wave equation. For this
purpose we define the coset generators as the generalized displacement
operators by means of the creation and annihilation operators $a$ and $a^{+}$%
. These operators are analogous to those corresponding to quadratic
Hamiltonians but the changement of sign for the generalized coordinate
introduces the imaginary frequency into the definition, (by analogy to the
generalized \textit{inverted} oscillator), namely 
\begin{align}
a & \; = \;\left( \frac{iz}{2\hslash\left\vert z\right\vert }\right)
^{1/2}\left( \sqrt{m\omega}\,q \;+ \;\frac{p}{\sqrt{m\omega}}\right) ,
\label{aad} \\
a^{+} & \; = \;\left( \frac{iz^{\ast}}{2\hslash\left\vert z\right\vert }%
\right) ^{1/2}\left( \sqrt{m\omega}\,q \;- \;\frac{p}{\sqrt{m\omega}}\right) 
\notag
\end{align}

Precisely, the change in the character of the frequency introduces the
global phase factor $e^{i\pi/4}.$

Consequently, the general displacement operator $D(\alpha )$ for any general
complex parameter $\alpha $ and $z$ takes the following form 
\begin{gather}
D\left( \alpha \right) S\left( z\right) \;=\;\exp {\left( \alpha
\,a^{+}-\alpha ^{\ast }a\right) }\;\times \;\exp {\frac{1}{2}\left(
za^{+2}-z^{\ast }a^{2}\right) }\;=\;  \label{des} \\
=\;\exp \left[ \left( \frac{i}{2\,\hslash \left\vert \,z\,\right\vert }\right)
^{1/2}\left( \sqrt{m\,\omega}\;\, d\,(\alpha, z)\, q \;-\; \frac{p}{\sqrt{m\,\omega}}\; d\,(\alpha, z)\,\right) \right] \exp \left[\, \frac{%
-i\,\left\vert \,z\,\right\vert }{2\,\hslash }\left( \,q\,p\,+\,p\,q\,\right)\,\right],
\notag
\end{gather} 
\be \label{d}
d \, (\alpha,\, z)\; \equiv \;\left( \;\alpha\; \sqrt{z^{\ast }}\;-\;\alpha ^{\ast }%
\sqrt{z}\;\right)
\ee
The displacement operator $D(\alpha )$ is a unitary operator. In the
coordinate representation $p\,=\,-\,i\;\hslash \;\partial_{q},$ the
displacement operator takes the form 
\begin{gather}
D\left( \alpha \right) S\left( z\right) \;=  \label{des1} \\
= \;\exp \left[ \left( \frac{i}{2\,\hslash \left\vert \,z\,\right\vert }\right)
^{1/2}\left( \sqrt{m\;\omega }\;d \, (\alpha,\, z)\;q \;+\;\frac{i\,\hslash }{\sqrt{m\;\omega }}\, d \, (\alpha,\, z)\,\frac{d}{dq}\,\right) \,\right] \exp %
\left[ \,-\,\left\vert\, z\,\right\vert \left( q\,\frac{d}{dq}\,+\,\frac{1}{2}%
\right) \right] ,
\end{gather}%
The operator $D(\alpha)$ Eq. (\ref{des1}) acts on the vacuum of the
inverted oscillator, namely%
\begin{equation}
\left\langle \;q\;|\;0\;\right\rangle _{inv-osc}\;=\;\left( \frac{im\omega }{%
\pi \hslash }\right)^{1/4}\exp \left( -\,\frac{im\omega }{2\hslash }%
\,q^{2}\,\right)  \label{fid}
\end{equation}

Therefore, we obtain the generalized coherent states with the following
form: 
\begin{align} \label{gcs}
\psi_{\alpha,z}\left( q\right) & \,=\,\left( \frac{im\omega}{\pi\hslash }%
\right)^{1/4}\exp\left[\,-\,\frac{1}{2}\left( \left\vert z\right\vert
\,+\,\left\vert \alpha\right\vert ^{2}e^{-2\left\vert z\right\vert }\,+\,%
\frac{e^{-\left\vert z\right\vert }}{\left\vert z\right\vert}\left(
\alpha^{2}z^{\ast}\cosh\left\vert z\right\vert
\,-\,\alpha^{\ast2}z\sinh\left\vert z\right\vert \,\right) \right) \right]
\cdot \\
& \exp\left[ -\,\frac{im\omega}{2\hslash}q^{2}\,+\,\sqrt{\frac{\,2im\omega }{%
\left\vert z\right\vert \hslash}}\left( \alpha\sqrt{z^{\ast}}\,\cosh\left\vert
z\right\vert \,-\,\alpha^{\ast}\sqrt{z}\,\sinh\left\vert z\right\vert
\,\right) e^{-\left\vert z\right\vert }\,q\,\right]
\end{align}
\\
These states are \textbf{squeezed} because the quantum uncertainty in
space and momentum coordinates is not equally distributed in the both
directions. In particular, we can test it putting $z = 0$, and we obtain the
coherent state for the \textbf{inverted} harmonic oscillator:

\begin{equation}\label{invho}
\psi_{\alpha}\left( q\right) _{inv-osc} \; = \; \left( \frac{i\,m\,\omega}{%
\pi\,\hslash}\right)^{1/4}e^{-\frac{1}{2}\left\vert \alpha\right \vert
^{2}}\,\exp\left( -\,\frac{i\,m\,\omega}{2\,\hslash}\,q^{2}\; + \;\sqrt{\frac{\,m\,\omega
}{\,2\,\hslash}}\;(\,1 +\,i)\,\alpha\, q\right)
\end{equation} \\
It is convenient to consider this type of coherent states as being based in
a Lie group $G$ with a unitary, irreducible representation $T$ \ acting on
some Hilbert space $\mathcal{H}$. If we take a fixed vector $\psi_{0}$ of $%
\mathcal{H}$, we define the coherent state system $\{T,\psi_{0}\}$ to be the
set of vectors $\psi \in\mathcal{H}$ such that $\psi = T(g)\,\psi_{0}$ for
some $g\in G$. Then, generalized coherent states are defined as the states $%
\left\vert \psi\right\rangle $ corresponding to these vectors in $\mathcal{H}$.

\medskip

We can see that in the definition of the coherent state of the
Klauder-Perelomov type, the general displacement operator contains, in the
exponential representation of the coset, a linear part in the annihilation and
creation operators and another quadratic part corresponding to the "squeezed"
sector, eg see Eq. (\ref{des}): The latter belongs in this representation to $%
Mp\left( 2\right) $. Therefore, at least for this purely squeezed part in
the $a^{2},a^{+2}$ representation the {\it complete} vacuum state is  
\begin{equation*}
\left(\, j \,+ \,ka^{+} \,\right) \,\left\langle \;q\;|\;0\;\right\rangle_{inv-osc}
\end{equation*}%
where ($j, k$) are constants determined by the normalization of states and the boundary conditions. This is a consequence of the action of the Metaplectic group which increases the spectrum of physical states: $\Delta
n = \pm 2$, because the complete states are spanded by both: the $\mathcal{H}_{1/4}$ states (eg. even $(2n)$ states ), and the $\mathcal{H}_{3/4}$ states (eg odd $(2n + 1)$ states). Thus, the lowest level  $(n \,= \; 0)$ is in $\mathcal{H}_{1/4}$,  while   in $\mathcal{H}_{3/4}$ it is $n = 1$.

Consequently, being the complete vacuum under the action of an element of $Mp\left( 2\right)$,  Eq. (\ref{fid}) would take as a wave function the precise form: %

\begin{equation}\label{vacMp2}
\left. \psi _{\text{\,vacuum}}\,\right\vert_{Mp\left( 2\right) }\;\;\rightarrow \;\,
\left( \;\frac{i\,m\,\omega }{\pi \,\hslash }\;\right) ^{1/4} e^{-\,\frac{i\,m\,\omega }{%
2\,\hslash }\,q^{2}}\left( 1\; + \;e^{\,i\,\pi /4}\,\sqrt{\frac{\,m\,\omega }{4\,\hslash }}%
\; q\,\right)
\end{equation} 
\subsection{Momentum  representation}

Analogously to the case of the representation of coordinates, the
generalized coherent states are calculated in the same way but taking into
account that in the moment representation $p$ remains the same but
$q = i\hslash\partial_{p}$ in all the operators, from those of
annihilation and creation, namely
\begin{align}
a_{p} &  \;\;=\;\; \left(  \frac{i\,z}{2\,\hslash\,\left\vert \,z \,\right\vert}\right)
^{1/2}\left(  \sqrt{m\omega}\;i\,\hslash\,\partial_{p}\;+\;\frac{p}{\sqrt{m\omega}%
}\right)  ,\\
a_{p}^{+} &  \;=\;\left(  \frac{i\,z^{\ast}}{2\,\hslash\,\left\vert \,z\,\right\vert
}\right)^{1/2}\left(  -\sqrt{m\omega}\;i\,\hslash\,\partial_{p}\;+\;\frac{p}%
{\sqrt{m\omega}}\right)  \nonumber
\end{align}
 as well as in the operators of displacement \, $D(\alpha)S\left(
z\right)$ Eq.(\ref{des1}) acts on the vacuum of the inverted
oscillator in the momentum representation, namely
\begin{equation}
\left\langle \;p\;|\;0\;\right\rangle _{inv-osc}\;=\;\left(  \frac{i}%
{\pi\hslash m\omega}\right)  ^{1/4}\exp\left(  -\,\frac{i\,p^{2}\,}{2\,\hslash
m\omega}\,\right)
\end{equation}
Therefore, we obtain the generalized coherent states with the
following form: 
\begin{align*}
\psi_{\alpha, z}\left(  p\right)   &  \,=\,\left(  \frac{i}{\pi\hslash m\omega
}\right)  ^{1/4}\exp\left[  \,-\,\frac{1}{2}\left(  \left\vert z\right\vert
\,+\,\left\vert \alpha\right\vert ^{2}e^{-2\left\vert z\right\vert }%
\,+\,\frac{e^{-\left\vert z\right\vert }}{\left\vert z\right\vert }\left(
\alpha^{2}z^{\ast}\cosh\left\vert z\right\vert \,-\,\alpha^{\ast2}%
z\sinh\left\vert z\right\vert \,\right)  \right)  \right]  \cdot\\
&  \exp\left[  -\,\frac{ip^{2}}{2\hslash m\omega}\,+\,\sqrt{\frac
{2i}{\left\vert z\right\vert \hslash m\omega}}\left(  \alpha\sqrt{z^{\ast}%
}\cosh\left\vert z\right\vert \,-\,\alpha^{\ast}\sqrt{z}\sinh\left\vert
z\right\vert \,\right)  e^{-\left\vert z\right\vert }\,p\right]
\end{align*}
\\
These states are squeezed because the quantum uncertainty in the
space and momentum coordinates is not equally distributed. In particular, we can test it putting $z=0$, and we
obtain  the coherent state for the inverted harmonic oscillator
in the $p$ representation:

\[
\psi_{\alpha}\left(  p\right)  _{inv-osc}\,=\,\,\left(  \frac{i}{\pi\hslash
m\omega}\right)^{1/4}e^{-\frac{1}{2}\left\vert \alpha\right\vert ^{2}}%
\exp\left(  -\,\frac{i\,p^{2}}{2\,\hslash m\,\omega}\,\,+\,\sqrt{\frac{1}{2\hslash m\omega}}\; (\,1\,+ \,i\,)\;\alpha \, p\right)
\]

\subsection{Wigner function quasiprobability}

As we have made mention, the inverted Hamiltonian is formally
obtainable from the standard harmonic oscillator by the change \, 
$\omega \,\rightarrow \,\pm \;i\;\omega$ \; and it corresponds to the Hamiltonian
of the harmonic oscillator with purely imaginary frequency. Therefore, this
replacement transforms the eigenfunctions of the harmonic oscillator into
generalized eigenvectors of the inverted harmonic oscillator, which, from the
spectral point of view, leads us to a discrete purely imaginary spectrum:\;
$E_{\,inv-osc}\,= \,\pm \,\,i\,E_{\,h-osc.}\,=\,i\,\hbar\,\omega\,\left(\,n+1/2\,\right)$.  Notice that the replacement $\omega \;\rightarrow\; \pm \; i\,\omega $ generates in the fiducial or fundamental states of the
inverted oscillator the following forms:  
\[
\left\langle \;q\;|\;0\;\right\rangle _{inv-osc}\; =\; \,\left( \, \frac{i\,m\omega
}{\pi\hslash}\,\right)^{1/4}\exp\left(  -\,\frac{i\,m\omega}{2\hslash}%
\,q^{2}\,\right) \qquad \text{and }
\]
\[
\widetilde{\left\langle
\;q\;|\;0\;\right\rangle }_{inv-osc}\; = \;\,\left( \frac{-i\,m\omega}{\pi\hslash
}\,\right)  ^{1/4}\exp\left(  \,\frac{i\,m\omega}{2\hslash}\,q^{2}\,\right).
\]
\;
Consequently, in this approach and in order to have functions to be
truly $L^{2}$, one must consider $\left\langle \widetilde
{\varphi}_{inv-osc.}^{\ast}\right\vert \left.  \varphi_{inv-osc}\right\rangle
$ to take the square norms, (e.g. a biorthonormalization condition).
Note that according to these symmetries, both for the oscillator states and
for the coherent states obtained here, it is fulfilled:
\[
\widetilde{\varphi}_{inv-osc}^{\ast}\left( q\right) \; = \;\varphi \left( q\right)  \text{ \ \ \ for the inverted oscillator}%
\]%
\[
\widetilde{\psi}_{\alpha}^{\ast}\left(  q\right)\;  = \; \psi_{\alpha}\left(
q\right)  \text{ \ \ \ for coherent states}%
\]
and this is a consequence of the underlying symmetry of \,$Mp(2)$ \,since the generator $T_{1}\,= \,\frac{1}{4}\left(qp\, + \,pq\right)\,= \,\frac{i}{4}\left(a^{+2}\,-\,a^{2}\right)$ \, is the one that produces the
rotation or mapping on the states of the harmonic oscillator, e.g.
\begin{align*}
\varphi_{inv-osc}\left(  q\right)   & \;= \;e^{-\frac{\pi}{2}\,T_{1}}\;\varphi
_{HO} \left(  q\right)  \\
\widetilde{\varphi}_{inv-osc}\left(q\right)   & \; = \;\,e^{\frac{\pi}{2}\,T_{1}%
}\;\varphi_{HO}\left(  q\right)
\end{align*}

Taking this fact and symmetries into account, the Wigner function quasiprobability will be defined as
\[
W\left(  q,p\right) \,\; =\; \,\int dv\; e^{ -\,i\, \frac{ p\,v}{\hbar}}\;\widetilde{\psi}_{\alpha
}^{\ast}\,\left(\,q -\frac{v}{2}\,\right) \, \psi_{\alpha} \,\left( \, q + \frac{v}%
{2}\,\right) 
\]
Obtaining explicitly: 

\[
W\left(  q, \,p\right) \,\; = \; \, \exp\left[ \,-\left( \, \frac{m\omega}{\hslash}\,q^{2}%
\;+\;\frac{p^{2}}{\hslash m\omega}\;-\;2\,\sqrt{\frac{m\omega}{\hslash}}\;\alpha \,
q\;+\;\left\vert \alpha\right\vert^{2}\,\right)\,\right]
\]
\newline
Similarly, in the momentum representation, the Wigner function will be defined as%
\[
W\left(  q,p\right)  =\int du\;e^{\frac{i\,q\,w}{\hbar}}\;\widetilde{\psi}_{\alpha
,0}^{\ast}\left(  p-\frac{u}{2}\right)\;  \psi_{\alpha,0}\left(p+\frac{u}%
{2}\right)
\]
Explicitly:
\[
W\left(q,p\right) \; = \;\exp\left[ -\left(  \frac{m\omega}{\hslash}\,q^{2} \;
-\;\frac{p^{2}}{\hslash m\omega}\;-\;2\,\sqrt{\frac{m\omega}{\hslash}}\;\alpha\;
p \:+ \;\left\vert \alpha\right\vert ^{2}\right)  \right]
\]
\newline
We can also consider the Wigner function for the pure squeezed case, namely%
\begin{equation}
\psi_{z}\left(  q\right)  \;=\;\left(  \frac{im\omega}{\pi\hslash}\right)
^{1/4}e^{-\,\frac{1}{2}\left\vert z\right\vert \,}\exp\left(  -\,\frac
{im\omega}{2\hslash}\,q^{2}\right)
\end{equation}
Again, the Wigner function in this case will be defined as%
\[
W_{sq}\left( q, p\right) \; = \;\int dv\;e^{-i\,\frac{p\,v}{\hbar}}\;\widetilde{\psi}%
_{z}^{\ast} \;\left( q -\frac{v}{2}\right)\; \psi_{z}\left(  q + \frac{v}%
{2}\right)
\]%
with the result:
\begin{equation}
W_{sq}\left(  q, \,p \right)  \; = \; e^{-\,\left\vert z\right\vert \,}\,\exp\left[
-\,\left(\frac{m\omega} {\hslash}\,q^{2}\,-\,\frac{p^{2}}{\hslash m\omega}\right) \,\right]
\end{equation} \;

which shows a purely Gaussian behaviour without the $\alpha-$shift tail.

 \medskip

As is it known the spectrum of the inverted oscillator gives rise to
complex generalized eigenvalues because from the point of view of the
potential this is unbounded from below. Here we saw that the fact of
considering the mapping given by the $T_1$ generator of the group Mp(2) allows to treat in
equal footing the respective solutions of the inverted oscillator and the standard harmonic one. 
It is worth mentioning that using quasi-hermiticity techniques \cite{oi} the problem can be analogously solved by considering a scaling operator proportional to $T_{1}$ . 
 
On the other hand,
solutions for the singular case (x-coordinates) lead to solutions of the
parabolic cylinder type. These solutions only reflect the symmetries of the
metaplectic vacuum, due that they are factorized, by means of hypergeometric
type functions, in an {\it even} part and an {\it odd} part corresponding precisely to the
two irreducible sectors of $Mp(2)$. It is useful to remember that the generators
of $Mp\left(2\right)$ are the following ones
\begin{align}
T_{1} &  = \;\frac{1}{4}\left( \, q\,p \,+ \,p\,q\,\right) \; = \;\frac{i}{4}\left(\,  a^{+2}%
- \,a^{2} \,\right)  ,\label{cr}\\
T_{2} & \; = \;\frac{1}{4}\left( \, p^{2}\,-\,q^{2}\,\right) \;\, =\;-\,\frac{1}{4}\left(\,
a^{+2}\,+ \,a^{2}\,\right)  ,\nonumber\\
T_{3} & \; = - \,\frac{1}{4}\left( \, p^{2} \,+\,q^{2}\,\right)  = \;- \,\frac{1}{4}\; \left(\,
a^{+}a \;+ \;a\,a^{+} \right)  \nonumber
\end{align}
With the following commutation relations,%
\[
\left[T_{3},\,T_{1}\right]\;= \; i\,T_{2},\qquad \left[  T_{3},\,T_{2}\right]\;  = \; -\,i\,T_{1}%
,\qquad \left[T_{1},\,T_{2}\right]\; = \;-\,i\,T_{3}%
\]
being $(q,\,p)$, alternatively ($a,\,a^{+}$), the variables of the standard harmonic
oscillator, as usual.

\medskip
 
In the following Section we describe the generalized coherent states in terms of the {\it Metaplectic group} 
MP(n), its metric constructed in  phase space and its associated
relativistic particle field equation.

\section{Metaplectic Group $MP(n)$, Algebraic
interpretation of the Metric and the Square Root Hamiltonian}

One of the basis of the dynamical description is the Hamiltonian or Lagrangian of the square root type,
that is, a non-local and non-linear operator in principle. This is because  the \textit{invariance under reparametrizations} as a Lagrangian and as an
associated Hamiltonian, generates the correct physical spectrum.
The essential guidelines of our approach here are based on the items specifically described in the sequel:
\begin{itemize}
\item{(i) The elementary distance function (positive square root of the line
element) is taken as the fundamental geometric object of the
space-time-matter structure, the geometric Lagrangian (functional action)
of the theory.}

\item{(ii) From (i) the geometric Hamiltonian is obtained in the usual way : this will be
the fundamental classical-quantum operator.}

\item{(iii) This universal Hamiltonian (square root Hamiltonian) contains a zero
moment $P_{0}$ characteristic of the complete phase space at the maximum level, from the point of view of the physical states. The 
inclusion of a $P_{0}$ prevents the arbitrary nullification of the
Hamiltonian, a fact that occurs  in the proper time system in
which the evolution coincides with the time coordinate: in this case time
"disappears" from the dynamic equations.

The method that we use to preliminarily expand the phase space to determine later here (via Hamilton equations) the physical role of $P_{0}$ is the
Lanczos method \cite{lanczos}, which is the most geometrically consistent and
mathematically simplest.}

\item{(iv) The Hamiltonian, when rewritten in differential form, defines a new
relativistic wave equation of second order and degree $1/2$. This can be
reinterpreted as a Dirac-Sudarshan type equation of positive energies and
internal variables (e.g. oscillator type variables), having a para-Bose or
para-Fermi interpretation of the solution-states of the system.}

\item{(v) The spectrum will be formed by states that are bilinear in fundamental
functions, which in the case of  $Mp\left(
2\right)$ are $f_{1/4}$ and 
 $f_{3/4}$ having a spin weight $s \,= \, 1/4$ and $3/4$ supported and connected by a vector
representation of the generators of $Mp\left( n\right)$, or those covered
by this, e.g. $SU\left(\, p, \, n-p \,\right),$ $SL(\,nR\,)$, etc. A
characteristic physical state of  $Mp\left( 2\right)$  is of the
form $\Phi _{\mu } \; = \;\left\langle \,s\,\right\vert $ $L_{\mu }\,\left\vert \,s^{\prime
}\,\right\rangle $ with $(s,\, s^{\prime} \,= \, 1/4, \;3/4 \,)$, \, and \, $L_{\mu } $ being the vector representation
of one of the generators of  $Mp\left( 2\right)$.}
\end{itemize}

\subsection{Mp(2), SU(1,1) and Sp(2)}

Following on the items (i) to (v) above, we use as a base the  
line element in a  $N \,= \,1 $ superspace with differential 
forms. Consequently, we extend our manifold to include fermionic
coordinates. Geometrically, we take as starting point the functional action
that describes the world-line (measure on a
superspace) of the superparticle as follows : 
\begin{equation}\label{S1}
S \;= \;\left(\, x,\,\theta ,\,\overline{\theta }%
\,\right) \;= \;- \,m \int_{\tau _{1}}^{\tau _{2}}\,d\tau \;\sqrt{\,\overset{\circ }{%
\omega_{\mu }}\overset{\circ }\,{\omega ^{\mu }}\; + \;{\mathbf{\gamma}\,\overset{.}%
{\theta }^{\alpha }\overset{.}{\theta }_{\alpha }\;-\;{\mathbf{\gamma}}^{\ast }\,%
\overset{.}{\overline{\theta }}^{\overset{.}{\alpha }}\overset{.}{\overline{%
\theta }}_{\overset{.}{\alpha }}}}  
\end{equation}
where \,$\overset{\circ }{\omega _{\mu }} = \overset{.}{x}_{\mu }\,- \,i\;(\,\overset{.}{%
\theta }\ \sigma _{\mu }\overline{\theta }\,-\,\theta \ \sigma _{\mu }\,\overset{.}%
{\overline{\theta }}\,)$,\, and the dot indicates derivative with respect to the
parameter $\tau $ as usual; the complex constant $\,\mathbf{\gamma}$ allows  generality to  
characterize the states and describing limiting cases

The above Lagrangian is constructed considering the line element (e.g.: the
measure, positive square root of the interval) of the non-degenerated
supermetric introduced in \cite{DiegoJMP} 
\begin{equation*}
ds^{2}\, = \, \omega^{\mu }\,\omega _{\mu }\,+\,{\mathbf{\gamma}}\,\omega^{\alpha }\,\omega
_{\alpha }-{\mathbf{\gamma}}^{\ast }\omega ^{\dot{\alpha}}\omega _{\dot{\alpha}},
\end{equation*}%
where the bosonic term and the Majorana bispinor compose a superspace $%
(1,3\,|\,1)$, with coordinates $(t,x^{i},\theta ^{\alpha },\bar{\theta}^{%
\dot{\alpha}})$, and where the Cartan forms of the supersymmetry group are
described by: $\omega _{\mu }\,=\,dx_{\mu } - i\,(\,d\theta \sigma _{\mu }\bar{\theta%
}\,-\,\theta \sigma _{\mu }d\bar{\theta}),\qquad \omega ^{\alpha } = d\theta
^{\alpha },\qquad \omega ^{\dot{\alpha}}\,=\,d\theta ^{\dot{\alpha}}$ (obeying
evident supertranslational invariance).

The {\it generalized momenta} from the geometric Lagrangian are computed
in the usual way:
\begin{equation}\label{P1}
\mathcal{P}_{\mu } \;= \;\partial L/\partial x^{\mu }\; = \;\left(\, m^{2}/L\,\right) 
\overset{\circ }{\omega _{\mu }}
\end{equation}%
\begin{equation} \label{P2}
\mathcal{P}_{\alpha } \;= \;\partial L/\overset{.}{\partial \theta ^{\alpha }}\;=\;i
\mathcal{P}_{\mu }\left( \sigma ^{\mu }\right)_{\alpha \overset{.}{\beta }}%
\overline{\theta }^{\overset{.}{\beta }} \, + \,\left(\, m^{2}\, \gamma/\,L\,\right) \overset{.}{%
\theta _{\alpha }}
\end{equation}%
\begin{equation} \label{P3}
\mathcal{P}_{\overset{.}{\alpha }}\;= \;\partial L/\overset{.}{\partial \overline{%
\theta }^{\overset{.}{\alpha }}} \; = \;i\,\mathcal{P}_{\mu }\theta ^{\alpha }\left(
\sigma^{\mu }\right)_{\alpha \overset{.}{\alpha }} \,- \,\left(\, m^{2}\,\gamma /L\,\right) 
\overset{.}{\overline{\theta }_{\overset{.}{\alpha }}} 
\end{equation}

We write them in a {\it canonical form} %
\begin{equation} \label{S}
\Pi _{\alpha }\;\;= \;\; \mathcal{P}_{\alpha }\,+\,i\ \mathcal{P}_{\mu }\,\left( \,\sigma
^{\mu }\,\right) _{\alpha \overset{.}{\beta }}\,\overline{\theta }^{\overset{.}{%
\beta }}
\end{equation}%
\begin{equation}
\Pi_{\overset{.}{\alpha }}\; \;= \;\; \mathcal{P}_{\overset{.}{\alpha }}\,- \, i\;\mathcal{P}%
_{\mu }\;\theta^{\alpha }\;\left(\, \sigma ^{\mu }\,\right)_{\alpha \overset{.}{%
\alpha }}
\end{equation}%
($\mathcal{P}_{\alpha }$ and $\ \mathcal{P}_{\mu }$ being defined from the
Lagrangian, as usual). Then, we start with the equation (which will become the wave equation) :
\begin{equation} \label{S}
\mathcal{S}\left[ \Psi \right] \;= \;\,\mathcal{H}_{s}\,\Psi \;= \;\sqrt{\,m^{2}\, - \,\mathcal{P}%
_{0}\,\mathcal{P}^{0}\,- \,\left(\, \mathcal{P}_{i}\,\mathcal{P}^{i}\,+ \,\frac{1}{\gamma}\,\Pi
^{\alpha }\,\Pi_{\alpha }\,- \,\frac{1}{\gamma^{\ast }}\,\Pi ^{\overset{.}{\alpha }}\,\Pi _{%
\overset{.}{\alpha }}\,\right) }\;\Psi   
\end{equation}

\medskip

As we have extended our manifold to include fermionic coordinates, it is
natural to extend also the concept of  a point particle trajectory to the
superspace. To do this, we take the coordinates $x\left( \tau \right) $, $%
\theta ^{\alpha }\left( \tau \right) $ and $\overline{\theta }^{\overset{.}{%
\alpha }}\left( \tau \right) $ depending on the evolution parameter $\tau .$

Consequently, there exist an algebraic interpretation of the
pseudo-differential operator (square root) in the case of an underlying
Metaplectic group structure Mp\thinspace $\left( n\right) $ : 
\begin{equation}
\sqrt{\;\mathcal{F}\;}\;\left\vert
\;\Psi \;\right\rangle\; \equiv\; \underset{}{\sqrt{\,m^{2}\,-\,\mathcal{P}_{0}\,\mathcal{P}^{0}\,-\,\left( \mathcal{P}_{i}\,
\mathcal{P}^{i}\; +\; \frac{1}{\gamma}\,\Pi^{\alpha }\,\Pi _{\alpha }\,-\,\frac{1}{\gamma^{\ast }}
\Pi^{\overset{.}{\alpha}}\, \Pi _{\overset{.}{\alpha }} \,\right) }\;\left\vert
\;\Psi\; \right\rangle \; = \;0}\label{sqrtH1}
\end{equation}%
\begin{equation}
\left\{ \; \left[ \;\mathcal{F} \;\, \right]
_{\beta }^{\alpha }\;\left( \Psi\, L_{\alpha }\;\right)\; \right\} \,\Psi ^{\beta } \;\equiv \;  
\left\{ \left[\, m^{2}\,-\,\mathcal{P}_{0}\,\mathcal{P}^{0}\,-\,\left( \mathcal{P}_{i}\,
\mathcal{P}^{i}\; + \;\frac{1}{\gamma}\,\Pi ^{\alpha }\,\Pi _{\alpha }\; -\; \frac{1}{\gamma^{\ast }}%
\Pi^{\overset{.}{\alpha }}\,\Pi_{\overset{.}{\alpha }}\,\right) \right]
_{\beta }^{\alpha }\left( \Psi L_{\alpha }\right) \right\} \Psi ^{\beta } = 0
\label{sqrtH2} 
\end{equation}

Then, both structures can be identified, e.g:
\be \label{Id}
\sqrt{\;\mathcal{F}\;}\;\left\vert
\;\Psi \;\right\rangle\;
\;\leftrightarrow \;\,\left\{ \; \left [ \;\mathcal{F} \; \right]
_{\beta }^{\alpha }\;\left( \Psi\, L_{\alpha }\;\right)\; \right\} \,\Psi ^{\beta },
\ee 
being the state $\Psi $ the square root of a spinor $\Phi $ (where the
"square root" Hamiltonian acts) such that it can be bilinearly defined as $%
\Phi \;= \;\Psi \;L_{\alpha }\;\Psi .$

\medskip The operability of the pseudo-differential "square root"
Hamiltonian can be clearly interpreted if it acts on the square root of the
physical states. In the case of the Metaplectic group, the square root of a
spinor certainly exist  \cite{sann}, \cite{majo},  \cite{dirac}, \cite{arv}
making the identification Eqs (\ref{sqrtH1})- (\ref{sqrtH2}) fully consistent both: from the relativistic
and \ group theoretical viewpoints. 

\medskip 

Is also possible to describe a complete multiplet spanning spins from $%
(\,0, \,\;1/2, \,\;1,\,\;3/2,\,\;2 \,).$ This is so because with the fundamental states and the allowed vectorial generators, the tower of states is finite and the states involved are  {\it all physical},  as it must be from the physical viewpoint.

\medskip

The choice of Eq.(\ref{S1}) as a functional
action in superspace is justified because from the point of view of
symmetries, it contains the largest symmetry algebra of the harmonic
oscillator with 3 quadratic generators in $a$ and $a^{+}$
($B_{0}$: even sector) and the two generators in the $B_{1}$: odd sector,  describing the superalgebra \,$Osp\,(1/2,\,R)$ 
 with its 5 generators.

\medskip

It is notable that in the general case, \,$Sp(2m)$\,  can be embedded somehow in a larger algebra as \,$Sp(2m)\;+\; R^{2m}$\; admitting
an Hermitian structure with respect to which it becomes the orthosymplectic superalgebra \, $Osp(2m,1)$. Consequently the metaplectic representation of \,$Sp(2m)$\,
extends to an irreducible representation (IR) of \,$Osp(2m,1)$\, which
can be realized in terms of the space $H$ of all holomorphic
functions \,$ h:C^{m} \,\rightarrow \,C/\int\left\vert h\left(  z\right)  \right\vert
^{2}e^{-\left\vert z\right\vert^{2}} d\lambda\left( z\right)  < \infty$ 
with $ \lambda\left( z\right)$ the Lebesgue measure on
$C^{m}$. The restriction of the $Mp\left(  n\right)$
representation to $Sp(2m)$,\, implies that the two irreducible sectors are
supported by the subspaces $H^{\pm}$ of $H$, where $H^{+}$ and $H^{-}$ are the spans (closed) of the set of functions $z^{n}\equiv\left( z_{1}^{n_{1}},....,\, z_{m}^{n_{m}}\right)$ with \,
 $n_{\theta}\,\in \,Z $,\;  $\left\vert n \right\vert = \sum n_{\theta}$, {\it even} and {\it odd} respectively.

\subsection{Geometrical Spinorial SL(2C) description of the Zitterbewegung}

Let us briefly analyze in an algebraic description, the origin of the quantum relativistic effects as the prolongated highly oscillations effect or so called 
"Zitterbewegung". There are two types of states: the basic
(non-observable) states and the observable physical states. The basic states
are coherent states corresponding to the double covering of the $SL(2C)$, eg
the Metaplectic group  \cite{DiegoP}, \cite{arv}
responsible for projecting the symmetries of the 6 dimensional $Mp(4)$ group
space to the 4 dimensional space-time by means of a bilinear combination of
the $Mp(4)$ generators. The supermultiplet solution for the geometric
lagrangian is given by
\begin{equation*} 
g_{ab}(0,\lambda) \;=\; \left\langle \psi_{\lambda}\left( t\right)\;
\right\vert\; L_{ab}\;\left\vert \;\psi_{\lambda}\left( t\right) \;\right\rangle
\end{equation*}
\begin{equation*}
g_{ab}\;(0,\lambda) \;= \; \exp{[\,A\;]} \;\exp{\left[\;\xi
\varrho\left(t\right)\;\right]} \;\chi_{f} \;\langle
\;\psi_{\lambda}(\,0\,)\;| \;\;\left( \,[c]{c}\;\right)_{ab}\,|\;\psi_{\lambda}(\,0\,)%
\;\rangle ,
\end{equation*}
\begin{equation} \label{A}
A \,(\, t\,) \; \; = \;\;-\,\left(  \frac{m}{\left\vert \gamma\right\vert
}\right)^{2}t^{2}\,+\,c_{1}\,t \, + \, c_{2} \; \; ,\qquad(c_{1}, \, c_{2})\;\in\;C 
\end{equation}

where we have written the corresponding indices for the simplest supermetric
state solution, being $L_{ab}$ the corresponding generators $\in Mp\left(
n\right) $, and $\chi_{f}$ coming from the odd generators of the big
covering group of the symmetries of the specific model. Considering for
simplicity the `square' solution for the three compactified dimensions  (spin $\lambda$ fixed, $\xi\equiv-\left( \overline{\xi}^{\overset{.}%
{\alpha}}-\xi^{\alpha}\right)$), the exponential even fermionic part is
given by: 
\begin{align}
\varrho\left( t\right) \;\equiv \;\overset{\circ}{\phi}_{\alpha} \;\left[\; \left(\,
\alpha\, e^{i\omega t/2}\right. \right. & \left. \left. +\right. \right.
\left. \left. \beta \,e^{-i\omega t/2}\,\right) \;- \;\left( \sigma^{0}\right)_{%
\overset{.}{\alpha}}^{\alpha}\,\left(\, \alpha \,e^{i\omega t/2}\;- \;\beta \, e^{-i\omega
t/2}\,\right)\, \right] \\
&  + \,\frac{2i}{\omega}\,\left[ \left(\, \sigma^{0}\,\right) _{\alpha}^{\overset{.}{\
\beta}}\ \overline{Z}_{\overset{.}{\beta}}\; + \;\left( \sigma^{0}\right)_{\ 
\overset{.}{\alpha}}^{\alpha}\ Z_{\alpha}\,\right]  \label{even}
\end{align}

$\overset{\circ}{\phi}_{\alpha}, \, Z_{\alpha},\,\overline{Z}_{\overset{.}{\beta}%
} $ being constant spinors, and $\alpha$ and $\beta$  $\mathbb{C}$-numbers
(the constant $c_{1}\in\mathbb{C}$ due to the obvious physical reasons and
the chirality restoration of the superfield solution.
By consistency, (and as in the string case), two geometric-physical options
are related to the orientability of the superspace trajectory: $
\alpha = \pm \beta$. We take without loss of generality $\alpha = + \beta$ \,
then, exactly, there are two possibilities:

\medskip

(i) The compact case which is associated to the small mass limit (or $\left\vert \gamma\right\vert >>1)$ :
\begin{equation}
\varrho\left( t\right) \;= \,\left( 
\begin{array}{c}
\overset{\circ}{\phi}_{\alpha}\cos\left(\,\omega\, t/2\,\right) \;+\;\frac{2}{\omega }\,
Z_{\alpha} \\ 
-\,\overset{\circ}{\overline{\phi}}_{\overset{\cdot}{\alpha}}\,\sin\,\left(\, \omega
t/2\,\right) \;-\;\frac{2}{\omega}\,\overline{Z}_{\overset{.}{\alpha}}%
\end{array}
\right)  \label{comp}
\end{equation}
(ii) And the non-compact case, which can be associated to 
 the imaginary frequency ($\,\omega \,\rightarrow \,i\,\omega\,$ generalized inverted oscillators) case:

\begin{equation}
\varrho \left( \,t\,\right) \;\; = \; \;\left( 
\begin{array}{c}
\overset{\circ}{\phi}\,\cosh\,\left(\, \omega\, t\,/2\,\right)\;+ \;\frac{2}{\omega }\;
Z_{\alpha} \\ 
-\;\overset{\circ}{\overline{\phi}}_{\overset{\cdot}{\alpha}}\,\sinh\,\left(\,
\omega\, t/2\,\right)\;- \;\frac{2}{\omega}\;\overline{Z}_{\overset{.}{\alpha}}%
\end{array}
\right)  \label{noncomp}
\end{equation}
\\
Obviously (in both cases), this solution represents a \textit{Majorana
fermion} where the $\mathbb{C}$ (or $hypercomplex$) symmetry wherever the
case) is inside the constant spinors.

\begin{figure}
[ptb]
\begin{center}
\includegraphics[scale=1.00]{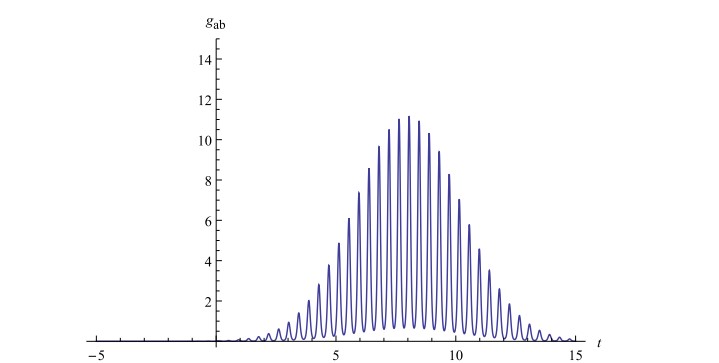}
\caption{Oscillation between the chiral and antichiral part of the bispinor
$\varrho(t)$, or \textit{Zitterbewegung,} for suitable values of the group
parameters. This oscillation reflects the underlying chiral and antichiral
quantum structure of the spacetime.}
\label{Fig.1}
\end{center}
\end{figure}

The spinorial even part of the superfield solution in the exponent becomes: 

\begin{equation}
\xi\,\varrho\left( t\right) \;= \;\theta^{\alpha}\;\left( \overset{\circ}{\phi }%
_{\alpha}\cos\left( \omega t/2\right) \;+ \;\frac{2}{\omega}\;Z_{\alpha}\;\right)\; - \;
\overline{\theta}^{\overset{\cdot}{\alpha} } \left( -\,\overset{\circ}{%
\overline{\phi}}_{\overset{\cdot}{\alpha}}\sin\left( \omega t/2\right) \;-\;%
\frac{2}{\omega}\;\overline{Z}_{\overset{.}{\alpha}}\;\right)  \label{67}
\end{equation}
We easily see that in the above expression there appear a type of continuous oscillation between the chiral and antichiral
part of the bispinor $\varrho(t)$, or \textit{%
Zitterbewegung} as shown  qualitatively in Fig.(\ref{Fig.1})
  for suitable values of the group parameters. 
This
oscillation reflects in our context the underlying chiral and antichiral quantum structure of
the spacetime. Thus, the
physical meaning of such a relativistic oscillation (Zitterbewegung) does
appear here as an underlying geometrical  supersymmetric effect, namely a kind of duality between supersymmetric and relativistic effects.

\medskip

In the next Section we provide more details about how the quantum
dynamics and  space-time structure emerge from this principle of symmetry.

\section{Relativistic Wave Equation and the Complete Hilbert Space} \label{completestates}

The importance of illustrating with this model based on the simplest
N = 1 supergroup is that it has a formal equivalence with known cases containing
the Poincare group, generally coming from symmetry breaking models with
minimum group manifolds SO(1,4 ), SO(2,3) as characteristic examples , $SUSY_{N=1} \;\sim\;SO(1,4).$ Recalling the
geometric Lagrangian constructed from the line element from the Maurer Cartan
forms induced via pullback (e.g. nonlinear realization for example) of a
fundamental symmetry group:
\begin{equation}
S \; = \;\int_{\tau_{1}}^{\tau_{2}}\; d\tau \; L\left( x\right)\; =
\;- \;m\,\int_{\tau_{1}}^{\tau_{2}}\,d\tau \,\sqrt{\,\omega_{AB}\;\omega^{AB}}
\end{equation}
$A, B \,= \,0 ,...., 5$.  The line element is based on the Cartan forms of the
symmetry group, for which it is induced and reflected in the geometric
Lagrangian. Consequently, for SO(1,4), for example, we have 10 that agree
with the number of generators of the group, as it must be, the indices of
the forms run from 0 to 4. If by some process, the symmetry is preferably
dynamically broken, the Cartan forms from the point of view of the algebra,
are divided into the 6 generators of SO(1,3) plus 4 generators of the Cartan
forms, namely $\rightarrow\;\omega_{AB}\;\rightarrow \;\omega_{\mu
\nu,}\;\omega_{\mu4}\;\sim\;\sqrt{\lambda}\,\theta_{\mu} \;(\,$Poincare - tetrad
fields\,), $\mu\,,\nu\, = \,0,...,3$%
\begin{align*}
S & \; = \;\int_{\tau_{1}}^{\tau_{2}}\;d\tau \; L\left( x\right) \; =\;-\;
m\int_{\tau_{1}}^{\tau_{2}}\,d\tau \sqrt{\,\omega_{\mu\nu}\;\omega^{\mu\nu}\; + \;%
\omega_{\mu4}\;\omega ^{\mu4}} 
\end{align*}

with \; $\omega_{\mu4}\;\omega ^{\mu4} 
 \;= \;\lambda\,\theta_{\mu}\,\theta^{\mu} $ . Following on the arguments given in the precedent paragraphs, 
we are going to see how wave equations for physical states emerge from the very spacetime
structure.

\subsection{Mp $\left( 2\right)$ - Coherent basic and bilinear states} \label{subcbbils}

Now we will demonstrate how the sector of the metaplectic group
becomes determinant in the problem of determining the geometric structure and
symmetries of the interplay between physical states and spacetime. To this
end, we know from the so-called positive energy equations, that
these types of equations should emerge.
We introduce the transformation (evolution-type
ansatz)
\[
\Phi_{\gamma}\left(  \,t\,\right)  \;\;=\;\;e^{\,[\, B\,\left(\, t \,\right)\; + \; p_{i}
x^{i} \; + \;\xi\,\varrho\;\left(\,  t\,\right) \;]\,}\;\Phi_{\gamma} \left(\,  0\,\right)
\]
Note that, in contrast to the case where only $\sigma
^{0} = I_{2}$\, comes into play, here we include the parameters $p_i$ in
order to generate the complete and less trivial matrix structure. Consequently,
\begin{gather} \label{p1}
\left[  \;\left\vert \,\gamma \,\right\vert ^{2}\left(\,  \partial_{0}^{2}%
\,-\,\partial_{i}^{2}\,\right)  \;+\;\frac{\Delta_{+} - \Delta_{-}}{4}\; + \;m^{2}\;\right]
^{1/2}\left\vert \;\Psi\;\right\rangle\; = \;0\\
\left\{\left[\,\left\vert \,\gamma \,\right\vert ^{2}\left(\,  \partial_{0}%
^{2}\,-\,\partial_{i}^{2}\,\right)  \;+\;\frac{\Delta_{+} - \Delta_{-}}{4}%
\;+\;m^{2}\;\right]  _{\beta}^{\alpha}\left\vert \;\Phi_{\alpha}\;\right\rangle
\;\right\}^{1/2}\;=\;0 \label{p2}
\end{gather} 
where $\Delta_{\pm}\;\equiv \;\left[\,  \partial_{\eta}\mp
\partial_{\xi}\;\pm \;i\ \sigma^{\mu}\partial_{\mu}\left(  \eta\;\pm\;\xi\, \,\right)\,\right]^{2}$ and we consider the equivalence at the level of
operators between the square root on the basic state of the
metaplectic $\left\vert \;\Psi\;\right\rangle $ \ defined as an
independent coherent state in each even or odd irreducible sector, and the radicand on the bilinear 
 $\Phi_{\alpha}$ = $\left\langle \;
\Psi\;\right\vert\; L_{\alpha}\;\left\vert \;\Psi\;\right\rangle$ \, 
written in the ket usual form : $\left\vert \Phi_{\alpha}\;\right\rangle
= \left(
\begin{array}
[c]{c}%
a\\
a^{+}%
\end{array}
\right)_{\alpha}\left\vert \;\Psi\;\right\rangle .$  
 Consequently, the sector 
 $B_{0}$\ (Bose)
generates the system %
\[
\left\vert \;\gamma\;\right\vert ^{2}\left(  \overset{\cdot\cdot}{B}%
\;+\;\overset{\cdot}{B}^{2}\;-\;p_{i}^{2}\right) \; + \;m^{2}\;=\;0
\]
where the function B is determined by
\[
B \;= \;\ln\;\left[\,  c_{2}\;\cos\, b\,(t) \; \right]
 , \qquad b \,(t) \;=  \; \sqrt{\;\frac{m^{2}}{\left\vert \,\gamma\,\right\vert
^{2}}\;-\;p_{i}^{2}}\;\,\left(t\;-\;t_{0}\right)  
\]
It is important to notice that in the general case $ B = \ln\;\left[\;
c_{2}\cos\,b\,(t) \;+\;c_{2}^{\prime}\sin\,
b \,(t) \; \right]$ \ we take 
 without losing generality $c_{2}^{\prime}%
= 0$  because we concentrate on the
Mp(2) part. It is easy to see that if $ c_{2}^{\prime} = ic_{2}$, the solution for $B$  is proportional to $\sqrt
{\frac{m^{2}}{\left\vert \,\gamma \,\right\vert^{2}}-p_{i}^{2}}\,\left(
t\,-\,t_{0} \right)$  and also to the Gaussian resolvent packet  with the factor $(\,\frac{m^{2}}{\left\vert
\gamma\right\vert^{2}} + p_{i}^{2}\,)$ instead of just $\frac{m^{2}%
}{\left\vert \,\gamma \,\right\vert^{2}}$.

The sector $B_{1}$\ (Fermi N = 1) gives us the equation
\[
\left\vert \,\gamma\,\right\vert^{2}\xi\left(  \overset{\cdot\cdot}{\;\rho
}\;+\;2\,\overset{\cdot}\,{\rho}\,\,\overset{\cdot}{B}\;\right) \; \;= \;\;0
\]
with a general solution of the form 
\begin{equation}
\varrho\left( \,t\,\right) \;=\;\frac{1}{c_{2}^{2}\;\sqrt{\;\frac{\;m^{2}}{\left\vert \,\gamma \,\right\vert
^{2}}\,-\,p_{i}^{2}}} \;\,\overset{\circ}{\phi}_{\alpha}\;  \left[\;
\alpha\;\tan\,b\,(t)  \;-\;\beta\;\left(  \sigma
^{0}\right) _{\ \overset{.}{\alpha}}^{\alpha}\ \sec \,b\,(t)\;\right] \label{71}%
\end{equation}
The two parts are not
independent (in chiral and antichiral zones). Therefore, the equation reduces finally to: 

\begin{equation}
\left(
\begin{array}
[c]{cc}%
-ip_{z}-\overset{\cdot}{B} \;\;\;\;& -ip_{x}-p_{y}\\
-ip_{x}+p_{y} \;\;\;\;& ip_{z}-\overset{\cdot}{B}%
\end{array}
\right)_{\beta}^{\alpha}\left\vert\; \Phi_{\alpha}\;\right\rangle \;\;= \;\; 0\label{maj}%
\end{equation}
\\
Knowing that: \textbf{(i) }$\left\vert \;\Phi_{\alpha} \;\right\rangle$  \;= \; $L_{\alpha
}\;\left\vert \;\Psi\;\right\rangle $ is the generator in vector
representation based on annihilation and creation operators, and that: \textbf{(ii)} It transforms as a spinor under the group SO(1,2), SU(1,1)
and Mp(2) (with the respective mappings between them), it is shown that $\left\vert \;\Psi \;\right\rangle$  is the coherent state formed by
two separate {\it even} and {\it odd} coherent states of the considered metaplectic group.
We explicitly have

\begin{equation} \label{majo1}
\left(
\begin{array}
[c]{cc}%
-ip_{z}-\overset{\cdot}{B} \;\;\;& -ip_{x}-p_{y}\\
-ip_{x}+p_{y} \;\;\;& ip_{z}-\overset{\cdot}{B}%
\end{array}
\right)_{\beta}^{\alpha}\left(
\begin{array}
[c]{c}%
a\\
a^{+}%
\end{array}
\right)_{\alpha}\left\vert \;\Psi\;\right\rangle \;\;= \;\;0 
\end{equation}
\\
which have exactly the same appearance as the equations of the type of
internal variables and positive energies of Majorana and Dirac for example.
This is easily seen by introducing the choice of parameters:  $p_{z} = -i\,\epsilon,\; p_{x} = 0, \; p_{y} = p$:

\begin{equation}\label{majo2}
\left(
\begin{array}
[c]{cc}%
\epsilon+\overset{\cdot}{B} \;& p\\
-p \;& -\epsilon+\overset{\cdot}{B}%
\end{array}
\right)  _{\beta}^{\alpha}\left(
\begin{array}
[c]{c}%
a\\
a^{+}%
\end{array}
\right)_{\alpha}\left\vert \;\Psi\;\right\rangle \;\;= \;\; 0
\end{equation}

Notice that $\overset{\cdot}{B}$ would take the formal role of
"mass" and the transformations are just of the squeezed type.

\subsection{The Mp(2)  Squeezed Vacuum and Physical States} \label{vacuumphys}

The displacement operator in the case of the vacuum squeezed is an element of Mp(2)
written in the respective variables of the canonical annihilation and
creation operators.
\begin{equation} \label{s}
S\left( \, \xi\,\right) \; = \;\exp\frac{1}{2}\left(\;  \xi^{\ast}a^{2}\;-\;\xi\,
a^{+2}\;\right)  \;\; \in \; \;Mp\left(  2\;\right)  %
\end{equation}
Seeing Eqs. (\ref{majo1}) and (\ref{majo2}) the relationship is shown
directly:

\begin{equation} \label{ss}
\left(
\begin{array}
[c]{c}%
a\\
a^{+}%
\end{array}
\right) \; \rightarrow \; S \left( \, \xi\,\right)  \left(
\begin{array}
[c]{c}%
a\\
a^{+}%
\end{array}
\right)  S^{-1}\left(\,\xi\,\right)  \;\,=\;\left(
\begin{array}
[c]{cc}%
\lambda & \mu\\
\mu^{\ast} & \lambda^{\ast}%
\end{array}
\right)  \left(
\begin{array}
[c]{c}%
a\\
a^{+}%
\end{array}
\right)  %
\end{equation} 
\\
From Eqs (\ref{s}) (\ref{ss}), we see that the dynamics of these
"square root" fields of $\Phi_{\gamma}$, in the particular
representation that we are interested in, is determined by considering these
fields as coherent states in the sense that they are eigenstates of $a^{2}%
$ via the action of the Mp$\left(  2\right) $ group that is of the
type:
\begin{align}
\left\vert \;\Psi_{1/4}\left(  0,\xi,q\right) \; \right\rangle  &  \;=\;\overset
{+\infty}{\underset{k=0}{\sum}}\;f_{2k}\left(  0,\xi\right)  \;\left\vert
\,2k\,\right\rangle \;=\;\overset{+\infty}{\underset{k=0}{\sum}}%
\;f_{2k}\left(\,0,\xi\,\right) \; \frac{\left(  a^{\dag}\right)^{2k}}%
{\sqrt{\,\left(\,2k\,\right)\,  !}}\;\left\vert \;0\;\right\rangle \label{psi1/4}\\
\left\vert \;\Psi_{3/4}\;\left(  \,0,\xi,q\right) \; \right\rangle  &
\; = \;\overset{+\infty}{\underset{k = 0}{\sum}}\;f_{2k+1}\;\left(\,  0,\xi
\,\right) \;\left\vert\, 2k+1 \,\right\rangle \;=\;\overset{+\infty}%
{\underset{k = 0}{\sum}}\;f_{2k+1}\left(  0,\xi\right) \; \frac{\left(
a^{\dagger}\right)^{2k+1}}{\sqrt{\,\left(\, 2k+1\,\right)\,  !}}\;\left\vert
\;0\;\right\rangle \nonumber
\end{align}

For simplicity, we will take all normalization and fermionic
dependence or possible fermionic realization, into the functions $f\left(
\xi\right)$. Explicitly, at $t=0$, the states are:
\begin{equation} \label{1/4-3/4} 
\begin{array}
[c]{c}%
\left\vert \,\Psi_{1/4} \;\left(  \;0,\,\xi,\,q\;\right) \; \right\rangle
\; \; = \;\;f\left(\,\xi\,\right)\;  \left\vert \;\alpha_{+}\;\right\rangle \\
\left\vert \,\Psi_{3/4} \,\left(  \,0,\,\xi,\,q\;\right) \; \right\rangle
\;\;=\;\;f\left(  \,\xi\,\right)\;  \left\vert \;\alpha_{-}\;\right\rangle
\end{array}
\end{equation}
where $\left\vert \alpha_{\pm}\right\rangle $\ are the CS
basic states in the subspaces \, $\lambda\;=\;\frac{1}{4}$ \,and \,
$\lambda \;=\;\frac{3}{4}$ \,  of the full Hilbert space. In other
words, the action of an element of $Mp\left(  2\right)\,$   keeps
them invariant (coherent), ensuring the irreducibility of
such subspace, e.g :
\[
\mathcal{H} \;\;\sim \;\;\left(
\begin{array}
[c]{cc}%
\mathcal{H}_{1/4} & \\
& \mathcal{H}_{3/4}%
\end{array}
\right)
\]
Consequently, the two symmetric and antisymmetric combinations $(\pm)$ of the two sets of states ${(1/4,\; 3/4 \,)}$  will span {\it all} the Hilbert space: $\mathcal{H}$:
\begin{equation} \label{completes}
\left\vert \,\Psi_{\pm} \;  \right\rangle \;\;= \;\; \left\vert \,\Psi_{1/4}\;  \right\rangle \;
\pm \;\left\vert \,\Psi_{3/4}  \; \right\rangle \;\;, \qquad
\left\vert
\; \pm \;\right\rangle \;\; = \;\; \left\vert
\; + \;\right\rangle \; \pm \;\left\vert \,-\,\right\rangle 
\end{equation}
 
And the general bilinear states are of
the type: 
$$\left\langle \;\pm\;\right\vert \;L_{\alpha}\;\left\vert \;\mp\;\right\rangle \quad 
\text{ and } \quad \left\langle \;\pm\;\right\vert \;\mathbb{L}_{\alpha}\;\left\vert\; \pm\;\right\rangle $$ where: $$L_{\alpha} = \left(
\begin{array}
[c]{c}%
\alpha\\
\alpha^{\ast}%
\end{array}
\right) , \;\; \mathbb{L}_{\alpha} =\left(
\begin{array}
[c]{c}%
a^{2}\\
\left(  a^{+}\right)^{2}%
\end{array}
\right)_{\alpha}; \;\;\;\;\;\left\vert \Psi_{1/4}\right\rangle =\left\vert
\,+\,\right\rangle, \;\;\left\vert \Psi_{3/4}\right\rangle = \left\vert \,-\,\right\rangle 
$$ 
For example, we have for the  states with the explicit form:  
\begin{equation} \label{75}
\Phi_{\alpha}\left(  t,\lambda\right)  \;=\;\left\langle \Psi_{\lambda}\left(
t\right)  \,\right\vert \mathbb{L}_{\alpha}\,\left\vert \,\Psi_{\lambda
}\left(  t\right)  \,\right\rangle \;=\;e^{\, A \,(\,t\,)}\;e^{\,\xi\,
\varrho\,\left(\,  t\,\right)}\,\left\langle \;\Psi_{\lambda}\left(0\right)
\;\right\vert \left(
\begin{array}
[c]{c}%
a^{2}\\
\left(a^{+}\right)^{2}%
\end{array}
\right)_{\alpha}\left\vert \Psi_{\lambda}\left(  0\right)  \;\right\rangle
\end{equation}%
\begin{equation} \label{76}
\Phi_{\alpha}\left(  t,\lambda\right)  \;=\;e^{\,A\, (\,t\,)}%
\,e^{\,\xi\,\varrho\,\left(\,  t\,\right)  }\,\left\vert \,f\left(  \xi\right)
\,\right\vert ^{2}\left(
\begin{array}
[c]{c}%
\alpha_{\lambda}^{2}\\
\alpha_{\lambda}^{\ast2}%
\end{array}
\right)_{\alpha}
\end{equation}
$\lambda$ being the helicity label or the spanned subspace, e.g. 
($ \pm$ ), and $ A (t) $ is given by 
\begin{equation} \label{A(t)}
A\;=\;-\,\left(\, \frac{m}{\left\vert\, \gamma \,\right\vert }\,\right)
^{2}t^{2}\,+\,c_{1}\,t\,+\,c_{2}; \;\qquad (\,c_{1},\;c_{2}\,)\;\in\;\mathbb{C} 
\end{equation}
The "square root" solution takes the
following form%
\begin{equation}\label{77}
\Psi_{\lambda} \, (\,t\,) \;=\;e^{\,\frac{1}{2}\, A\,(\,t\,) }
\; e^{\frac{\xi\,\varrho\left(  t\right) }{2}}\;\left\vert \; f\left(  \xi\right)
\,\right\vert \left(
\begin{array}
[c]{c}%
\alpha\\
\alpha^{\ast}%
\end{array}
\right)_{\lambda}
\end{equation}
where $\lambda\, = \,(1/4,\;3/4)$.  Notice the difference with the
case of the Heisenberg-Weyl realization for the states $\Psi$ :
\begin{equation}  \label{78}
\left\vert \,\Psi\,\right\rangle \;=\;\frac{f\left(  \,\xi\,\right)  }%
{2}\;\left(  \,\left\vert \;\alpha_{+}\;\right\rangle \; + \;\left\vert \;\alpha
_{-}\;\right\rangle \,\right)  \; = \;f\left(  \xi\right)  \left\vert
\;\alpha\;\right\rangle
\end{equation}
where, the linear combination of the states $\left\vert
\alpha_{+}\right\rangle$  and  $\left\vert \alpha_{-}\right\rangle$
span now the full Hilbert space, being for this CS basis $\lambda = \frac{1}{2}$. The "square" states
at $t = 0$  are  %
\begin{align}
\Phi_{\alpha}\left(0\right)   &  \; = \; \left\langle \;\Psi\left(  0\right)\;
\right\vert \; L_{\alpha}\; \left\vert \;\Psi\left(  0\right)  \;\right\rangle
\; = \; f^{\ast}\left(  \xi\right) \, f\left(  \xi\right)  \left(
\begin{array}
[c]{c}%
\alpha\\
\alpha^{\ast}%
\end{array}
\right)_{\alpha}
\end{align}
The square state and the obtained square root state at time $t$ are:
\begin{equation}  \label{81}
\Phi_{\gamma}\left(  t\right)  \; = \;e^{\,A }\; e^{\xi\varrho\left(  t\right)  }\;\;\left\vert \; f\left( \, \xi\,\right)\;
\right\vert^{2}\left(
\begin{array}
[c]{c}%
\alpha\\
\alpha^{\ast}%
\end{array}
\right)_{\alpha}, \quad 
\Psi\left(  t\right)  \; = \; e^{\frac{1}{2}A} \; e^{\,\frac{\xi\,\varrho\,\left(\,t\,\right) }{2}%
}\;\left\vert \;f\left(\,  \xi\,\right) \;\right\vert \left(
\begin{array}
[c]{c}%
\alpha^{1/2}\\
\alpha^{\ast1/2}%
\end{array}
\right)  
\end{equation}

Let us discuss the obtained results:

{\bf(i)} \, We can see that the algebra, carrying the topological
information of the group manifold, is "mapped" over the spinors solutions
through the eigenvalues $\alpha$  and $\alpha^{\ast}$ from
the dynamical viewpoint. The constants in the exponential functions of the
Gaussian type in the solutions come from the action of a unitary operator over the respective
coherent basic states in each Irreducible representation.

\medskip

{\bf(ii)} \, The $Osp\left(  1/2,\mathbb{R}\right)$ supergroup
allows a metaplectic representation containing the complete superalgebra in
functions of a single complex variable $z$ exactly coinciding with the example
treated here: it contains $SU(1,1)$ as subgroup which can lead or explain
the fermionic factors of the type $[\,\exp\,(\;\frac{\xi\,\varrho\,\left(
t\right)}{2}\;)\,]\times \left\vert \,f\left(\xi\right)\,\right\vert $ in
the solutions.

\medskip

{\bf(iii)} \, The $K_{\pm}$ and $K_{0}$ generators operate
over the Bose states ($B_{{0}}$ sector). The $B_{{1}}$ 
sector of the algebra given by $a$ and $a^{+}$ operates over
the fermionic part. In this case, the coherent and squeezed states that can be constructed
are eigenstates of the displacement and squeezed operators respectively (as in
the standard case) but they cannot minimize simoultaneously the dispersion of
the quadratic Casimir operator, such that they are not minimum uncertainty
states. This is so because the only states which minimize the Schrodinger uncertainty
relation are those obtained by applying the displacement or squeezed
operator on the lowest normalized state.

\medskip

{\bf(iv)} \, Geometrically, in the description of any physical system through
SU(1,1) coherent states (CS) or squeezed states (SS), the orbits will appear as the intersections of constant-energy surfaces with one sheet of a two sheeted
hyperboloid - the curved phase space of $SU(1,1)$ or Lobachevsky
plane - in the space of averaged algebra generators. The group containing
$SU(1,1)$ as subgroup linear and bilinear functions of the algebra
generators, can factorize operators as the Hamiltonian or the Casimir operator
(when averaged with respect to the group CS or SS): this  defines corresponding
curves in the averaged algebra space. If the exact dynamics is confined to the
$SU(1,1)$ hyperboloid, the validity of the Ehrenfest's theorem for the coherent or squeezed states implies that it necessarily coincides with the variational motion that
derives from the Euler- Lagrange equations for the Lagrangian
\[
\mathcal{L}\;=\;\left\langle \;z\;\right\vert \;i\;\frac{\widehat{\partial}%
}{\partial t}-\widehat{H}\;\left\vert \;z\;\right\rangle ,
\]
that will be different if $\left\vert \,z \,\right\rangle \;=\;\left\vert \,
\alpha\,\right\rangle$ or $\left\vert \,z\;\right\rangle \;=\;\left\vert\;
\alpha_{\pm}\;\right\rangle $, as it is evident.

\subsection{Discrete representation} \label{subdiscrete}

Eq.(\ref{77}) describes a  standard coherent state (eigenstate of the operator ($a$) as a linear
combination of two states belonging to $\mathcal{H}_{1/4}$ and $\mathcal{H}%
_{3/4}$ respectively, (which are two independent coherent states as eigenstates of
($a^{2}$).  The corresponding Metaplectic vacuum as fiducial
vector of the physical system is:
\begin{equation}
\left\vert \;z_{0}\; \right\rangle_{Mp(2)} \; = \; \mathcal{M} \;(1\;+\;\mathcal{M}^{2}\,a^{+})\;\left\vert\;
0\;\right\rangle \label{mpv}%
\end{equation}
\begin{equation}
\mathcal{M} \;\;\equiv\;\; [\;\left\vert m^{2} \;-\;\epsilon^{2}\;\right\vert \;+ \;p^{2} \;\text{sign}
 \,(\,\epsilon^{2} \,- \,m^{2}\,) \;]^{1/4}
\end{equation}
Notice that this vacuum
is not singular at $m\,\rightarrow \,\epsilon$ but is analytically continued into
the complex plane where it is defined. Then, the solution for Eq.(\ref{majo2}) is
the following:
\begin{equation} \label{Sz0}
\left\vert \;\Psi \;\right\rangle_{Mp(2)}\;\;\equiv \;\; S \left(\,t,\,A,\,p,\epsilon\,\right)\;
\left\vert \;z_{0}\;\right\rangle_{Mp(2)}
\end{equation}

\begin{equation*}
\left\vert \; \Psi \;\right\rangle_{Mp(2)}\; = \; \left[\,1 \;+\;\frac{p^{2}\,\text{sign}\;\mathcal{E}}{\vert \,\mathcal{E}\,\vert}\,\right]
^{1/4}e^{\;\frac{p/2}{(m\,+\,\epsilon)}\,(a^+)^{2}} \left[ \; 1 + \;\left(\;1\; +\;\frac{p^{2} \,\text{
sign}\;\mathcal{E}
}{\vert \,\mathcal{E}\,\vert}\,\right)^{1/2}a^{+}\; \right] \,\left\vert \;0\;\right\rangle \label{css}%
\end{equation*}
\begin{equation}
 \mathcal{E} \;\; \equiv \; \;\epsilon^{2} \;- \;(\overset{\cdot}{A})^{2}  
\end{equation}

being \,$S \left(\,t,A,p,\epsilon\,\right)
\in Mp\left( 2\right)$ the operator
Eq.(\ref{s}) for the set of parameters and functions in Eq. (\ref{majo2}). The
total solution of the system  Eqs.(\ref{p1}),(\ref{p2}) for these 
parameters  being $\;\mathcal{G}\,\left\vert \Psi\right\rangle_{Mp(2)}$ with
$\mathcal{G} \; \equiv \; e^{\,\left(\, A \,+ \,\xi\,\rho\,\right)(\,t,\,m,\,p,\,\epsilon
\,)}\,e^{\,(\,p\,y -i\,\epsilon\, z\,)}$ .

\medskip

The Bargmann representation of $\mathcal{H}$ associates an entire analytic
function $f\left(z\right)$ of a complex variable $z$, with each
vector $\left\vert \varphi\right\rangle  \in \mathcal{H}$ in the following
manner:%
\begin{align}
\left\vert \varphi\right\rangle  &  \in\mathcal{H} \; \rightarrow \; f\left(
z\right) \;\; = \;\; \underset{n\,=\,0}{\overset{\infty}{\sum}} \left\langle \,n \,\right.
\left\vert \,\varphi \,\right\rangle \frac{z^{n}}{\sqrt{\,n\,!}} \label{25}\\
\left\langle \varphi\right.  \left\vert \,\varphi\,\right\rangle \;  &
\equiv \;\left\vert \left\vert \, \varphi \,\right\vert \right\vert^{2} \; \;= \;\;\underset
{n \,=\, 0}{\overset{\infty}{\sum}}\;\left\vert \left\langle\, n \,\right.  \left\vert \,
\varphi \,\right\rangle \,\right\vert^{2} \; = \;\int\frac{d^{2}z}{\pi}\,e^{-\left\vert z\right\vert ^{2}}\,\left\vert f\left(
z\right) \, \right\vert^{2} \label{27}%
\end{align}
where the integration is over the entire complex plane. The above association
can be compactly written in terms of the normalized coherent states. Consequently:

\medskip

{\bf(i)}\, The $\mathcal{H}_{1/4}$ states occupy the sector even of the full Hilbert
space $\mathcal{H}$ and we describe them as:
\begin{equation} \label{36}
f^{\left(  + \right)  }\left(  z,\omega\right)  \; = \;\left(  1 \;- \;\left\vert
\omega\right\vert ^{2}\right)  ^{1/4}e^{\omega z^{2}/2}
 \; = \;\left( 1 \;-\; \left\vert \omega\right\vert ^{2}\right)^{1/4}\underset
{n \,=\, 0,\, 1,\, 2,..}{\sum}\frac{\left(  \omega/2\right)^{2n}}{(2n)!}\;z^{2n} 
\end{equation}
Then, in the vector representation we have:%
\begin{equation}
\left\vert \;\Psi^{\left(  + \right)  }\left(  \omega\right)  \;\right\rangle
\;= \;\left(1 -\left\vert \omega\right\vert ^{2}\right)^{1/4}\underset
{n \,=\, 0, 1, 2,..}{\sum}\frac{\left(  \omega/2\right)^{2n}}{\sqrt{(2n)\,!}}\;\left\vert
2n\right\rangle \label{37}%
\end{equation}
Consequently, the $\mathcal{H}_{1/4}$  (or {\it even}) number representation is obtained as:%
\begin{equation}
\left\langle \;2n \;\right.  \left\vert \; \Psi^{\left( + \right)  }\left(
\omega\right)\; \right\rangle \;  = \;\left(  1 \;-\;\left\vert \omega \right\vert
^{2}\right)^{1/4}\;\frac{\left( \omega/2 \right)^{2n}}{\sqrt{(2n)\,!}} \,, \qquad \; 
\left\langle 2n + 1\right.  \left\vert \Psi^{\left(+\right)}\left(
\omega\right)  \right\rangle \; \equiv \;0 
\end{equation}

\bigskip 

{\bf(ii)}\, The $\mathcal{H}_{3/4}$ states occupy the odd sector of the full
Hilbert space $\mathcal{H}$ and we  similarly describe them as for $\mathcal{H}_{1/4}$ :
\begin{equation}
f^{\left(  -\right)  }\left(  z,\omega\right) \;=\; \left(  1\;-\;\left\vert
\omega\right\vert^{2}\right)^{3/4} z\;e^{\,\omega\, z^{2}/2}\;
  =\;\left( 1\; - \;\left\vert \omega\right\vert ^{2}\right)  ^{3/4}\underset
{n\; =\; 0,\, 1,\, 2,..}{\sum}\frac{\left(  \omega/2\right)^{2n +1}}{(2n +1)!}\; z^{2n+1} 
\end{equation}
and the vector representation is:%
\begin{equation}
\left\vert \,\Psi^{\left(  -\right)  }\left(  \omega\right)  \,\right\rangle\;
= \;\left(  1\;- \;\left\vert \omega\right\vert ^{2}\right)  ^{3/4}\underset
{n\;= \;0,\,1,\,2,..}{\sum}\frac{\left(  \omega/2\right)^{2n + 1}}{\sqrt{\left(
2n+1\right)\, !}}\;\left\vert\, 2n+1\,\right\rangle \label{41}%
\end{equation}
The $\mathcal{H}_{3/4}$ (or  {\it odd }) number representation is consequently:%
\begin{equation}
\left\langle \,2n+1\,\right.  \left\vert \,\Psi^{\left(-\right)}\left(
\omega\right) \, \right\rangle \;=\;\left(  1\;- \;\left\vert \omega\right\vert
^{2}\right)^{3/4}\;\frac{\left(\omega/2\right)^{2n + 1}} {\sqrt{
(2n+1)\,!}} \,,\qquad \;
\left\langle \,2n\right.  \left\vert \,\Psi^{\left(  -\right)  }\left(
\omega\right) \, \right\rangle  \;\equiv \;0 \label{43}%
\end{equation}

{\bf(iii)} \,The full Hilbert space, defined by the direct sum $\mathcal{H}%
=\mathcal{H}_{1/4} \oplus \mathcal{H}_{3/4},$ is the following:%
\begin{align}
f\left(  z,\omega\right)  \; &  = \; f^{\left(  +\right)  }\left(  z,\omega\right) \;
+ \; f^{\left(  -\right)  }\left(  z,\omega\right) \label{44}\\
f\left(  z,\omega\right)  \; &  =  \;\left(  1\;- \;\left\vert \omega\right\vert^{2}\right)  ^{1/4}\underset
{n\,=\,\,0,\,1,\,2,..}{\sum}\frac{\left(  \omega/2\right)^{2n}}{(2n)!}\,z^{2n}\left[\,
1\;+\;\frac{\left( \, 1\;-\;\left\vert \omega\right\vert ^{2}\,\right)^{1/2}}{(2n+1)} \,z\;\right]  \label{45}%
\end{align}
Then, in complete analogy with their {\it even} and {\it odd} subspaces, the corresponding
states are:%
\begin{align}
\Psi\left(  \omega\right)   & \; = \; \Psi^{\left(  + \right)  }\left(
\omega\right) \;\; + \;\;\Psi^{\left( - \right)  }\left(  \omega\right) \label{46}\\
\Psi\left( \omega\right)   & \; = \;\;\left(  1\;-\;\left\vert \omega\right\vert ^{2}\right)^{1/4}\underset
{n \;= \;0, \,1,\, 2,..}{\sum}\frac{\left(  \omega/2\right)^{2n}}{\sqrt{(2n)!}} \,\left[\,
1 \; + \; \frac{\left(  1\;- \;\left\vert \omega\right\vert ^{2}\right)^{1/2}}{(2n+1)}\;a^{+}\,\right]
\left\vert \,2n \,\right\rangle \label{47}%
\end{align}%
\begin{equation}
\left\langle\, n \,\right.  \left\vert \,\Psi\left(  \omega\right)  \,\right\rangle \; = \;
\left\{
\begin{array}
[c]{c}%
\left(  1\;-\;\left\vert \omega\right\vert ^{2}\right)^{1/4}\;\frac{\left(
\omega/2\right)^{2n}}{(2n)\,!}\;\sqrt{\,2n\,!} \qquad \text{ even states\,}  \\
\\
\left( 1\;-\;\left\vert \omega\right\vert ^{2}\right)  ^{3/4}\;\frac{\left(
\omega/2\right)^{2n + 1}}{(2n + 1)\,!}\;\sqrt{\,\left(2n + 1\right)\,  !}  \qquad \text{odd states }
\end{array}
\right.  \label{48}%
\end{equation}
\\
where the link between the physical observables and the group parameters is
given by the following expression (measure):
\begin{equation} \label{49}
\left(  1 \; + \;\frac{p^{2} \,\text{sign}\left( \, \epsilon^{2}\,- \,m^{2}\,\right)}{\left\vert\,
m^{2}\,-\,\epsilon^{2}\,\right\vert }\right)^{1/4}\rightarrow \;\; \left( \, 1\;- \;\left\vert
\omega\right\vert ^{2}\;\right)^{1/4} 
\end{equation}

\begin{figure}
[ptb]
\begin{center}
\includegraphics[scale=1.00]{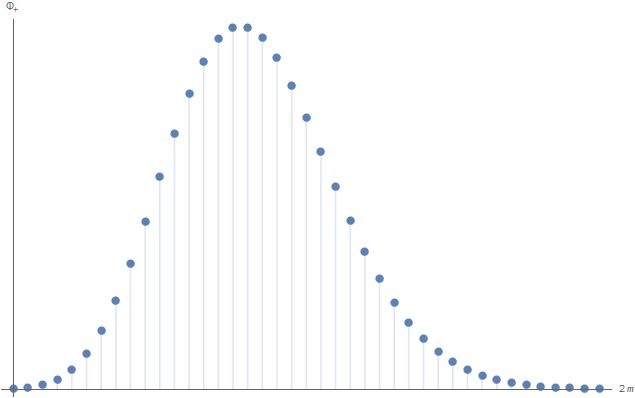}
\caption{The $\mathcal{H}_{1/4}$ discrete (number representation) states
occupy the even sector of the full Hilbert space $\mathcal{H}$. This Irreducible representation \ of the
 Mp $\left(2\right) $ group is not dense  
 (in a topological sense) but it
contains the ground state $\left\vert\, 0 \, \right\rangle$.}
\label{Fig.2}
\end{center}
\end{figure}

\begin{figure}
[ptb]
\begin{center}
\includegraphics[scale=1.00]{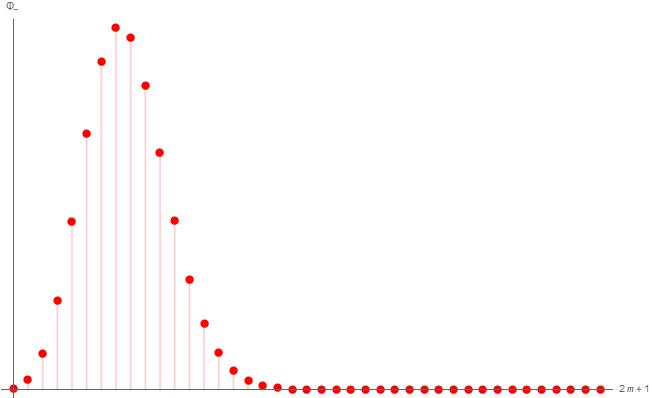}
\caption{The $\mathcal{H}_{3/4}$ discrete (number representation) states
occupy the sector odd of the full Hilbert space $\mathcal{H}$. This Irreducible representation of 
Mp $\left(2\right)$ is not dense (in a topological sense) but its lower or
fundamental state is the first excited  state 
 $\left\vert \, 1 \,\right\rangle $.} 
\label{Fig.3}
\end{center}
\end{figure}

Fig.(\ref{Fig.2}) and Fig.(\ref{Fig.3}) display the discrete spectra in the number
representation of the coherent states in $\mathcal{H}_{1/4}$ (even $n$) and
$\mathcal{H}_{3/4}$ (odd $n$).

\medskip

{\bf The Limit ${\bf\epsilon\rightarrow {\it m}: }$}

\medskip

This is precisely the limit $ \left\vert \,\omega\,\right\vert
^{2} \rightarrow 1$, which from the point of view of the Metaplectic analysis
corresponds to the edge of the complex disc. As we could easily see, the
state solutions span the full spectrum corresponding to $\mathcal{H}$. What
happens is that in the limit $\epsilon\rightarrow m$ the density of states
corresponding to $\mathcal{H}_{1/4}$ is greater than that of the odd states
belonging to $\mathcal{H}_{3/4}$.  It is for this reason that the states belonging
to $\mathcal{H}_{1/4}$, will survive in this limit. 

\section{Quantum Space-Time : de Sitter and Black Hole Coherent States} \label{deSBH} 

\subsection{Quantum Space-Time} 

We restrict in the sequel to the purely bosonic space-time and consider the $(X, T)$ quantum space and time dimensions which are relevant to the
quantum space-time structure. The remaining spatial transverse dimensions $%
X_{\bot} $ are not considered here as fully quantum non-commuting
coordinates. Notice that although the transverse spatial dimensions ${\bot}$
have zero commutators they can fluctuate. This corresponds to quantize the
two-dimensional space-time surface which is relevant to determine the
light-cone structure. This is enough for considering the novel features
arising in the global quantum space-time and the \textit{quantum light cone}. 

\medskip
 
The relevant  quantum space-time $(X,T)$ structure is described essentially by a quantum
inverted oscillator type algebra with discrete hyperbolic levels $(X^2 - T^2)_n = (2n +
1), \, n = 0, 1, 2, ...$. The zero point  energy $(n = 0)$ being the Planck energy
level. The truly quantum gravity (trans-Planckian) vacuum in the quantum space time is delimitated
by the four quantum  hyperbolae $ X^2 - T^2 = \pm 1$ (in Planck units) of the Planck scale $(n = 0)$ level. This is precisely a constant curvature \textbf{de Sitter} vacuum.  

\medskip

The {\bf de Sitter} space-time can be described as a (\textit{inverted}, ie with imaginary
frequency) harmonic oscillator, the \textit{oscillator constant and length}
being \cite{NSPRD2021},\cite{Sanchez3}: 
\begin{equation}  \label{oscdeS}
\kappa_{osc} \; = \;H^2, \qquad H\; = \; \sqrt{ \frac{(\,8\pi\,G\,\Lambda)}{3%
}} \;=\; c\,/\,l_{osc}
\end{equation}

The \textit{oscillator length} $l_{osc}$ is classically the Hubble radius, the Hubble
constant $H = \kappa$ being the \text{surface~gravity}, as the black hole
surface gravity is the inverse of (twice) the black hole radius.

Interestingly, the description of de Sitter space-time as an (inverted, classical and quantum) harmonic oscillator
derives from three results: 

{\bf(i)}\, From the Einstein Equations on the one hand  , \cite{NSPRD2021}, \cite{NSPRD2023}, \cite%
{dVSiebert}, \cite{dVSanchez}, 

{\bf(ii)}\, From the de Sitter geometrical description on the other hand: an hyperboloid embedded in 
flat Minkowski space-time with one more spatial dimension : 
\begin{equation}  \label{dS}
- \;T^2\; +\; X^2 \;+ \;X_i^2 \;+ \;Z^2 \;=\; L_{QG}^2
\end{equation}
\begin{equation}
L_{QG}\; = \;(\,L_Q \;+ \;L_G\,)\; = \;l_P\;\left(\,\frac{H}{h_P} \;+ \;\frac{h_P%
}{H}\,\right), 
\end{equation}
$L_{QG}$ is the complete length allowing to describe both the classical,
semiclassical and quantum (trans-Planckian) gravity domains, $l_P$ the constant Planck length:
\begin{equation}
L_Q\; = \;l_P^2 \,/\,L_G,\qquad  l_P \;= \; (\,2 \,G\,\hbar\, /\,c^3 \,)^{1/2}, \qquad h_P \,= \,{c} \,/\, {l_P}
\end{equation}
{\bf(iii)} \, From the hyperbolic quantum space-time structure 
which delimitates a purely quantum trans-Planckian central region of constant curvature , \cite{Sanchez2}, \cite{NSPRD2021}
\cite{NSPRD2023}
\medskip

In the \textbf{Anti-de Sitter} space-time, the description is the same but with 
$- T^2 + X^2 + X_i^2 + Z^2 \;= \;-\; L_{QG}^2$, \;
and therefore Anti- de Sitter
background is associated to a real frequency \textbf{(non inverted)}
harmonic oscillator. Also, the propagation of fields and linearized
perturbations in the de Sitter vacuum all satisfy equations which are like
the \textit{inverted} oscillator equations, \cite{GuthPi}, \cite{Albrecht}, 
\cite{SanchezNPB1987}, or the normal oscillator equations in Anti de Sitter space-time.

\medskip

In the (Schwarzschild) \textbf{black hole} space-time: (quantum interior constant curvature vacuum; semiclassical and classical exterior regions), the physical magnitudes as the oscillator
constant $ H^2$ and the typical oscillator length $l_{osc}$ are related to the black
hole mass $M$: 
\begin{equation}
H\;= \;{c}\,/\,{l_{osc}} \;=\; h_P \left(\,\frac{m_P}{M}\,\right), \qquad
\Lambda \; = \;{\lambda_P} \left(\,\frac{m_P}{M}\,\right)^2,
\quad \lambda_P \; = \; 3\, h_P^2 \,/\,c^4
\end{equation}

Classical space-time regions or regimes 
are described by the low values of $\Lambda$ and of the gravitational density  $\rho_G$,  and the large classical gravitational sizes $L_G >> l_P$: 

\begin{equation}
L_G \;\; = \;\; l_P\;\sqrt{\frac{\; \lambda_P}{\;\Lambda}}\;\; = \; \;l_P\; \left(\,\frac{M%
}{m_P}\,\right)
\end{equation}
 Truly Quantum gravitational regimes, eg in the trans-Planckian domaine of very small sub-Planckian sizes, very high quantum density $\rho_Q$ and very high vacuum values $\Lambda_Q$:
\begin{equation}
L_Q \,\;= \; \,l_P\; \sqrt{\,\frac{\,\Lambda} {\,\lambda_P} } \,\;=\;\,l_P\;\left(\,\frac{m_P}{M}%
\,\right), \qquad \Lambda_Q \;= \;\frac{\lambda_P^2}{\Lambda}
\end{equation}
Consistently,
the \textit{high} value of the classical/semiclassical gravitational entropy $S_G$  is
equal (in Planck units) to such high $\Lambda_Q$ value. This is clearly explicitated by the following classical-quantum gravity duality relations in  this context:
\begin{equation}  \label{LambdaHvalue1}
\frac{\rho_G}{\rho_P} \; = \;\left (\frac{l_P}{L_{G}}\right)^2 \; = \;\left(%
\frac{m_P}{M}\right)^2 \; = \;\left(\frac{S_Q}{s_P}\right)
\end{equation}
\begin{equation}  \label{LambdaQvalue2}
\frac{\rho_Q}{\rho_P} \; = \; \left(\frac{l_P}{\Lambda}\right) \; = \; \left(%
\frac{M}{m_P}\right)^2 \; = \;\left(\frac{S_G}{s_P}\right)
\end{equation}
\begin{equation*}
\rho_P \; = \;{3 \; h_P^2}\,/\,{8 \pi G}, \qquad s_P \; = \; \pi \kappa_B
\end{equation*}

The last r.h.s. of Eqs.(\ref{LambdaHvalue1})-(\ref{LambdaQvalue2}) show the
link to the gravitational \textit{entropy}:  quantum gravitational entropy $S_Q$ and
classical/semiclassical $S_G$ entropy. (This last is the Bekenstein-Hawking-Gibbons entropy 
 \cite{GibbHawkEntropy1} -  \cite{bekenstein1981}).   Lower case magnitudes with subscript $P$ denote the corresponding Planck scale fundamental constant magnitudes.

\medskip

The external BH region is precisely a \textit{classical gravity
dilute vacuum}, which $(\Lambda,\rho_G)_{BH}$ values in the present universe cannot be larger than the
observed very low values of $(\Lambda,\rho_G)$ Refs 
\cite{Riess} to \cite{Riess2022b}. Their quantum duals provide an upper bound to the high
values $(\Lambda_Q,\, \rho_Q)$ in the quantum central BH vacuum region
as determined by Eqs. (\ref{LambdaHvalue1})-(\ref{LambdaQvalue2})

\subsection{QST deS and BH. Minimal uncertainty and Mp(2) vacuum} 

For the quantum space-time (QST) {\bf de Sitter} states, the oscillator parameters entering
in the coherent states and their representations Section \ref{coherentstates} are the following:

\medskip

As discussed above, de Sitter space-time is described by an {\it inverted}
oscillator with oscillator length $l_{osc}\,= \, c / H $, and the generic
quantum coherent states built in Section \ref{coherentstates}, have in particular the inverted
oscillator length $l_{osc} \,=\, \sqrt{\hbar/m\omega}$.

The {\bf de Sitter} quantum space-time coherent states are described by the states
 Eqs (\ref{des}) - (\ref {des1}), Eq.\ref{gcs}) with the  corresponding  oscillator constant given by: 
\begin{equation}  \label{l-deS}
l_{osc\;dS}^{-2} \;= \; \left(\,\frac{ m \omega}{\hbar}\,\right)_{dS} \; =
\; H^2 \;= \; \frac{\Lambda}{3}
\end{equation}

In the (Schwarzschild) \textbf{black hole} space-time, the physical
magnitudes as the oscillator constant and the oscillator length are related
to the black hole mass $M$: 
\begin{equation}\label{l-BH}
l_{osc\; BH}^{-2} \;= \; \left(\,\frac{ m \omega}{\hbar}\,\right)_{BH} \; =
\;\, l_P^{-2} \,\left(\,\frac {m_P}{M} \,\right)^2
\end{equation}
\begin{equation*}
l_P \;= \; (\,2 \,G\,\hbar\, /\,c^3 \,)^{1/2}, \, \qquad h_P \;= \; c\,/\,l_P
\end{equation*}

The complete length \,$L_{QG}$\,in Eq.(\ref{dS}) covers both the classical, semiclassical and quantum (trans- Planckian) gravity domains. Quantum space-time derives from the quantum non commutative space and momentum (phase space) operators with the mapping of momentum into time, Refs (  \cite{Sanchez2}, \cite{NSPRD2021}, \cite{NSPRD2023}, \cite{Sanchez2019}). As a consequence, quantum space-time described by coherent states have minimal and equally distributed uncertainty:  $ \Delta X \;\Delta T \; = \; \hbar\,/\,2$
\begin{equation}  \label{deltaxt2}
(\,\Delta X \,)^2\; = \;\left(\frac{\hbar}{2\,m \,\omega} \right), \qquad
\quad (\,\Delta T\,)^2\; = \;\left(\frac{\hbar\,m\,\omega} {2}\right)
\end{equation}%
Therefore, coherent states of quantum de Sitter space-time have the spatial and temporal uncertainty: 
\begin{equation}\label{deltaxdeS}
(\,\Delta X \,){\,^2}_{\;deS} \; = \;\;\left(\frac{\hbar}{2\,m \,\omega}
\right)_{deS} \; = \; \;\frac{\hbar \,}{2\, H^2} 
\end{equation}
\begin{equation}\label{deltatdeS}
(\,\Delta T\,)_{\;deS}^{\,2}\; = \; \;\frac{\hbar\, H^{\,2}}{2\,}
\end{equation}
And for the Black Hole coherent states, the quantum uncertainty in space and time is: 
\begin{equation} \label{deltaxBH}
(\,\Delta X \,)_{\,BH}^2 \; \; = \; \;\left( \frac{\hbar}{2\,m\,\omega}\right)_{BH}  = \;
\;\;\frac{l_P^{\;2}}{2} \;\left(\,\frac{M}{m_P} \,\right)^2
\end{equation}
\begin{equation}\label{deltatBH}
(\,\Delta T\,)_{\;BH}^2\; \, = \; \;\frac{t_P^{\;2}}{2}\; \left(\;\frac{m_P}{M} \;\right)^2
\end{equation}

\medskip

($l_P$ and $t_P$ being the Planck length and time). 
The de Sitter and black hole coherent states derive from the explicit expressions Eqs (\ref{des}) - (\ref {des1}), Eq.(\ref{gcs}) with the respective (deS) and (BH) physical 
magnitudes given by Eqs (\ref{l-deS}) and Eqs (\ref{l-BH}). In particular, the quantum metaplectic Mp $\left(\,2\,\right)$ vacuum  is given by:
\begin{equation} \label{vacMp2deS}
\left(\,\left. \psi _{\text{\,vacuum}}\,\right\vert_{MP(2)}\,\right)_{deS}\;\;=\;\, 
\left( \;\frac{i\,}{\pi \,}\;\right) ^{1/4} \; \sqrt{\, H\,}\; e^{-\,\frac{i}{2}\,\left(\,H
\,X\right)^2} \,\left[\,\frac{\,H\,X }{2\,\sqrt{\,2}}\;(\,1\; + i\,)\; + \; \frac{\;i\; H\,X }{2\,\sqrt{\,2}}\; \right]
\end{equation}
\begin{equation*}  \label{vacMp2BH}
\left(\,\left. \psi _{\text{\,vacuum}}\,\right\vert_{MP(2)}\,\right)_{BH}\;\; = \;\, 
\left( \;\frac{i\,}{\pi \,}\;\right) ^{1/4} \; \sqrt{\,2 \,\mathcal{K} \,}\;\; e^{-\,\frac{i}{2}\,\left(\,2\,\mathcal{K}\,X\, \right)^2\,} \left[\; \frac{\,\mathcal{K} \,X }{\sqrt{\,2}}\;(\,1\; + \;i\,) \;  + \; \frac{i\;\mathcal{K} \,X} {\sqrt{\,2}}\; \right]
\end{equation*} 
where:  
\be \label{K} 
\mathcal{K}\;\; = \;\; 1 \,/ (\,2\,R_{BH}\,)\;\; =  1 \,/ (\,4\,G\,M\,)
\ee 
Both vacuum states are expressed in terms of the surface gravity ($H$ or $\mathcal{K}$ ) respectively, or similarly in terms of the de Sitter or BH radius. Both states are {\it totally  regular}, as it must be for quantum space-time. For $X >> R_{BH}, \;(\,  R_{BH}$ being the BH radius),  and asymptotically for very large $ X $, the quantum coherent state consistently accompasses the quantum space classicalization, as such exterior BH regions are semiclassical and classical. We discuss below the excited ($\alpha$) states.

\subsection{Continuum and Discrete  deS and BH Coherent states}

Quantum space-time de Sitter and black hole coherent states follow from Eqs  
(\ref{des}) - (\ref {des1}) and Eq (\ref{gcs}) with the  physical magnitudes and uncertainty relations Eqs (\ref{deltaxdeS} -  (\ref{deltatBH}). The quantum space-time $de S$ coherent states have the following expressions : 

\begin{equation}\label{csdeSs}
\psi_{\alpha} \left( X\right) _{deS} \; = \; \, 
\left( \;\frac{i\,}{\pi \,}\;\right) ^{1/4} \; \sqrt{\, H\,}\; e^{-\frac{1}{2}\,\left\vert \,\alpha\,\right \vert
^{2}}\;\exp {\,\left[\;\frac{\alpha \, H\,X}{\sqrt{\,2}} \;( 1 \;+\;i)\; - \; \frac{i\, H^2 \,X^{2}}{2}\;\right]}
\end{equation}

\begin{equation}\label{csdeSt}
\psi_{\alpha} \left( T\right)_{deS} \; = \; \, 
\left( \;\frac{i\,}{\pi \,}\;\right)^{1/4} \; \frac{1}{\sqrt{\,\hslash \, H\,}}\; e^{-\frac{1}{2}\,\left\vert \,\alpha\,\right \vert
^{2}}\;\exp {\,\left[\; \frac{\,\alpha\, T}{\sqrt{\,2}\;\, \hslash\, H} (\,1 \; + \; i\,) \; - \; \frac{ i \,T^{2}}{\,2\; \hslash^2 \,H^2} \,\right]}
\end{equation} 
 \\
 A  similar coherent state expression holds for the BH space-time with the corresponding  $BH$ factor $ 2\,\mathcal{K}$  instead of $H$, being  $ \mathcal{K}$ the surface gravity Eq. (\ref{K}).

$\alpha$ is the complex constant number, eigenvalue of the displacement operator $D\,(\,\alpha \,)$, which characterizes the coherent state excitations (displacement from the vacuum), and their continuum spectrum. 

\medskip

{\bf (i)}  The quantum space-time coherent states Eqs (\ref{csdeSs}) - (\ref{csdeSt}) clearly display an exponential de Sitter expansion term $\,\alpha\, T /(\, \sqrt{2} \,\hslash \,H \,)$ plus a phase (linear and quadratic) in $[(\,\alpha\, T /\, (\sqrt{2} \,\hslash \,H\,) \,]$, which is simply $ \,T \,/(\,2\,\sqrt{2}\, \pi\, T_H $, \, $T_H$ being the Hawking de Sitter Temperature.
The quantum space-time exhibits an accelerated expansion plus quantum oscillations of the same sign (linear term), and of different sign ( quadratic term). The presence of these oscillations is a new feature of quantum space-time.

\medskip
 
{\bf (ii)} The continuum $(\alpha)$-coherent states Eqs.(\ref{csdeSs}) - (\ref{csdeSt}) describe semi-classical, (or semi-quantum), space- time regimes  and in agreement with the space-time 
described by quantum oscillators. Quantum discrete space-time becames more and more
continuous for large $n$  in agreement with its description by continuum coherent states. Consistently, the continuum coherent states are characterized by the Hawking temperature which is a semiclassical (or semiquantum) temperature.

{\bf (iii)}\, We see that: 
\begin{equation*} 
\left\vert \,\Psi_{\alpha} \left (\, X \,\right)_{deS}\,\right\vert^{2} \;\,
- \; \;\left\vert \,\Psi_{\alpha} \left (\,T\,\right)_{deS}\,\right\vert^{2} \;\; = 
\end{equation*}
 \begin{equation}\label{hyperbs-t}
 = \;\; \frac{1}{\sqrt{\,\pi}}\;\exp^{-\,\left\vert \,\alpha\,\right\vert^{2}}    \,\left\vert\;H\; \exp{\,[\;c\,(\alpha)\; H \,X\;]} \;  - \; \frac{1}{\,\hslash\,H}\;\exp{\,[\;c\,(\alpha)\, \frac{\;T }{\hslash\, H} \;]}\;\right\vert\;\;
 \end{equation}
$$
c\,(\alpha) \;\;=\;\;\sqrt{\,2}\;\,
(\, \text{Re}\,\alpha \;-\; \text{Im}\,\alpha\;)
$$
which reflects the quantum hyperbolic space-time structure. Let us define:
$$
\mathcal{R}_{\alpha} \,(\,X,\, T \,)^2 \;\equiv \;\;\left\vert \,\Psi_{\alpha} \left (\, X \,\right)_{deS}\,\right\vert^{2} \;\,
 - \; \left\vert \,\Psi_{\alpha} \left (\,T\,\right)_{deS}\,\right\vert^{2} 
 $$   
 
{\bf(i)} For $\alpha \;=\; 0$  : 
\be 
\mathcal{R}_{0} \,(\,X,\,T \,)^2 \;\;= \;\;
\;\frac{1}{\sqrt{\,\pi}}\;\left\vert\, H \;\,-\;\,  \frac{1}{\,\hslash\,H}\;\right\vert 
\ee
which can be also expressed in terms of the quantum uncertainties:
\be
\mathcal{R}_{0} \,(\,X,\, T \,)^{\,2}\;\;=\;\;
\sqrt{\,\frac{2}{\pi\,\hslash}} \;\, \left \vert\; \Delta\,{T} \;- \; \frac{\Delta\,X}{\hslash} \;\right \vert
\ee
{\bf(ii)} For: 
\be
 \mathcal{R}_{\alpha} \,(\,X, \,T)^2 \;\;= \;\;0 \quad
\rightarrow \;\;\left\vert \,\Psi_{\alpha} \left (\, X \,\right)_{deS}\,\right\vert \;\,=
 \;\;\pm \; \left\vert \,\Psi_{\alpha} \left (\,T\,\right)_{deS}\,\right\vert 
 \ee
Clearly, $\alpha = 0$, corresponds to $\,H \,=\, 1 / \sqrt{\,\hslash}\,$, that is the Planck scale.  We see the power  of coherent states in describing space-time and even accounting for the Planck scale,  at which: 
\be
\Delta\,X = {\hslash \,/\,\sqrt{\, 2}} \,, \; \qquad \; \Delta\,T = 1\,/\,\sqrt{\, 2} \qquad \quad \text{( Planck scale)}
\ee
Obviously, for coherent states it satisfies $\Delta\,X \, \Delta\,T = \hslash \,/\,2$.  

\bigskip

For the {\bf squeezed} states, particularly interesting is the {\bf Wigner} quasi probability function, which have here the following expression:
\begin{equation}
W_{sq}\left(  X \,T \right)_{deS}  \; = \;\,\exp\left[
-\,\left(\,H^2\,X^{2}\,-\,\frac{T^{2}}{\hslash^2 \, H^2}\,\right) \,\right]
\end{equation}
This clearly shows the hyperbolic structure of quantum space-time. The characteristic light-cone structure is manifest here because there is no any $\alpha$-deformation in this case.

\medskip

Fig. 4 displays the space-time squeezed state Wigner function and its light- cone hyperbolic structure. 

\begin{figure}
[ptb]
\begin{center}
\includegraphics[scale=0.80] {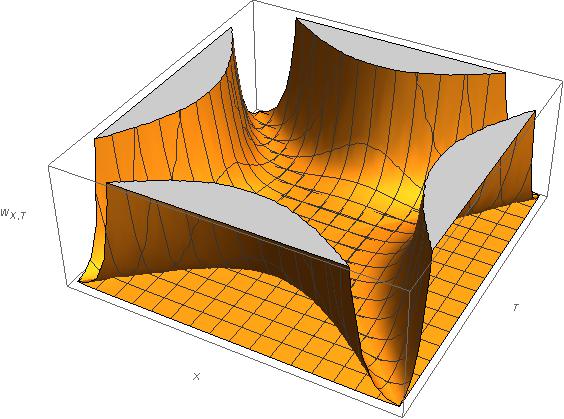}
\caption{The squeezed quasi-probability Wigner function $ W (X,T)$ of quantum space-time. $ W(X,T)$ clearly shows the hyperbolic light-cone space-time structure and with symmetric form. For the coherent states, $W_{\alpha} (X, T)$  endowes the hyperbolic structure but with a linear $\alpha$ tail deformation in $ X$ or  $T$.} 
\label{Fig.4}
\end{center}
\end{figure}

{\bf (iii)} The {\bf discrete} quantum space-time (Planckian and trans-Planckian) regimes
are described by {\bf discrete} states, eg. the discrete coherent states of subsection  
 (\ref{subdiscrete}). The discrete spectrum of
these states describes the different quantum space-time excitation levels,
the less excited $(\text{fundamental}, n = 0)$ level corresponding to the
Planck scale, (the {\it crossing} or transition scale). Interestingly, as seen in Section (\ref{completestates}) and (\ref{subdiscrete}) the Metaplectic group states with its {\it both} sectors and discrete representations, $\vert\, 2\,n \,\rangle$ \,and \, $\vert \,2\,n+1 \,\rangle$, {\it even} and  {\it odd} states, fully {\it cover} the {\it complete} Hilbert space $\mathcal{H}$  \be
\mathcal{H}\;\; =\;\; \mathcal{H}_{\,(\, + \,)} \;\; \oplus\;\;\mathcal{H}_{\,(\,- \,)} 
\ee
The $(\,\pm \,)$ symmetric and antisymmetric sum of the two kind ({\it even} and {\it odd})  
states  provides the {\it complete} covering of the Hilbert space and of the space-time mapped from it:  
\begin{equation}
\Psi \left(n\right)  \;\; = \;\; \Psi^{\left(+ \right)} \left(
2n\right) \;\; + \;\;\Psi^{ \left(- \right)} \left( 2n + 1\right) 
\end{equation}
where \;$ \Psi^{\left(+ \right)} $ \, and \,$\Psi^{ \left(- \right)} $ are obtained from Eqs (\ref{46})\,-\,(\ref{48}).
For de Sitter space, both sets of states are given by: 
\begin{equation} \label{stateH1}
\Psi^{\left(+ \right)} \left(
2n\right)_{deS}\; = \;
\left\vert\,1\;-\;H^{4}\,\right\vert  ^{1/4}\;\;\frac{\left(
H^2/2 \right)^{2n}}{\sqrt{\,(2n)\,!}}
\end{equation}
\begin{equation} \label{stateH2}
\Psi^{ \left(- \right)} \left( 2n + 1 \right)_{deS} \; = \;  \left\vert \,1 \;-\; H^{4}\,\right\vert^{3/4} \;\, \frac{\left(\, H^2 /2 \,\right)^{2n + 1}} {\sqrt{\,\left(2n + 1\right)\,!} }
\end{equation}
where we take into account that in the fully quantum trans-Planckian de Sitter phase  \cite{NSPRD2021}: the quantum  $ H$ is $H \, > \,1 $ and thus the analytic  covering in this phase. In addition, the quantum discrete levels of $ H $  are \cite{NSPRD2021} : $H_{Q\,n}\; = \;\sqrt{\,2 n}$\, \,({\it even} levels),\; and \; $H_{Q\,n} \; = \;\sqrt{\,2n+1}$, ({\it odd} levels), which leads :
\begin{equation} \label{discreteH1}
\Psi^{\left(+ \right)} \left(
2n \right)_{deS}\; = \;
\left \vert\; 1 \;-\; 4n^2 \; \right \vert^{1/4}\;\;\frac{\,(\,2n \,)^{2n}}{ 2^{2n} \,\sqrt{\,(2n)\,!}}
\end{equation}
\begin{equation} \label{discreteH2}
\Psi^{ \left(- \right)} \left( 2n + 1 \right)_{deS} \; = \;  [\;4n\,(n+1)\;]^{3/4} \;\,\frac{\left(\, 2n+1 \,\right)^{2n + 1}} { 2^{2n+1}\,\sqrt{\,\left(2n + 1\right)\,!} }
\end{equation}

It is worth mentioning that independently of this Mp(n) coherent state framework, we obtained in Refs 
 \cite{Sanchez2}, \cite{NSPRD2021}, \cite{Sanchez2019},  similar discrete levels in terms of   
 the global cart $X$, or the local ones $x$ constructed from the 
 global (complete) classical - quantum  duality including gravity \cite{Sanchez2019}. In such levels, the two kind of sectors and their global $(\pm)$ covering do appear, which reflects somekind of relation between the $ Mp(n)$ symmetry and classical-quantum duality: 
\begin{equation}  \label{Xn}
X_n \;= \; \sqrt{\,2\,n + 1\,},\quad \text{or}
\quad x_{n \pm} \;= \;[\;\, X_n \;\pm \;\sqrt{\, X_n^{2} \;- \;1 }\;\,],\quad
 \;\;n = 0,\, 1,\, 2\,, ....
\end{equation}

The condition $X_n^2 \geq 1$ simply corresponds to the whole spectrum $%
n\geq 0$ : 
\begin{equation}
x_{n \pm} \;= \; [\;\sqrt{\;2n +1 }\; \pm \;\sqrt{\;2n}\;\;]
\end{equation}
\begin{equation*}
x_{n = 0}\;(+) \;= \;x_{n = 0}\;(-) \;= \; 1: \;\mbox{Planck scale},
\end{equation*}
which complete {\it all} the levels. The $(\pm)$ branches consistently reflect : 
\begin{itemize}
\item{The classical- quantum duality properties of the {\it global} space-time.}
\item{ The two $\sqrt{\,(2n+1)}$ and  $\sqrt{\,2n}$, {\it even} and {\it odd}  (local) sectors. Each symmetric or antisymmetric sum is necessary to cover the whole manifold.  
 The corresponding $(\,\pm\,)$ global states are complete, CPT and unitary, the levels $n = 0, 1, 2, ....,$ cover the whole Hilbert space $\mathcal{H}\; =\;\mathcal{H}_{\,(\, + \,)} \;\; \oplus\;\;\mathcal{H}_{\,(\,- \,)}$ and all space-time regimes.}
\item{ The total $n$ states range over \textit{all} scales from
the lowest excited levels to the highest excited ones covering the two dual branches $(+)$ and $(-)$ or Hilbert space sectors.}
\end{itemize}

\section{Imaginary Time. Coherent States of  Quantum Gravitational Instantons}\label{s-tinstantons}

Taking imaginary time $T = i \mathcal{T} $, $t = i \tau$, yields to the elliptic (or circular) structure of space-time and of the phase space, eg this corresponds in particular to the  normal (non inverted) oscillator description.   
That is to say, quantum space-imaginary time {\bf instantons} correspond to the real frequency quantum oscillators of phase space. They describe in particular, quantum tunneling effects between different states or different vaccua, or different phase (space) regions. Besides being saddle points in an euclidean quantum gravity path integral, they can describe thermal features if the imaginary time endowes periodicity. 
 
 \bigskip
 
In the classical (non-quantum) BH space-time, the identification $T = i\, \mathcal{T} $, $t = i \,\tau$, transforms the hyperbolic
space-time structure into a circular structure:  The
classical horizon $X = \pm_, T \; $ collapses to the origin  $X = \pm \,\mathcal{T} = 0$. In the classical (non-quantum) BH \textit{instanton}, the  interior \textit{is cutted}, no horizon, and no central curvature 
singularity, does appear: The \textbf{classical} BH
instanton is \textit{regular} but  \textit{not complete}: The interior
BH region is \textbf{not} covered by the classical instanton.

\bigskip

In the complete quantum BH space-time,  the  quantum hyperbole 
$ (\,X^2 - T^2 = l_P^2\,)$ \, replace 
 the characteristic lines 
 due to the non-zero \,$[X,T]$ \,conmutators, and in the corresponding quantum BH instanton the horizon does not collapse to the origin  but to the Planck scale circle $ (\,X^2 + 
\mathcal{T}^2 = \,l_P^2 \,)$. The complete \textit{quantum} BH instanton includes the usual
classical/semiclassical BH instanton for radius larger than the
Planck length, plus a \textit{new central} highly dense \textit{quantum core}
of Planck length radius and high constant and \textit{finite} curvature corresponding to the \textit{black-hole interior}, Ref \cite{NSPRD2023} which is \textit{%
absent} in the classical BH instanton.

\medskip

Particularly interesting here is the {\bf Wigner quasiprobability function} for the  {squeezed} states,  which  for the BH  have the following expression:

\begin{equation}\label{wignerBH}
W_{sq}\left( \, X \,\mathcal{T} \,\right)_{BH}  \;\, = \;\,2\,\,\exp\left[
-\,\left(\,4 \,\mathcal{K}^2 \,X^{2}\; + \;\frac{\mathcal{T}^{2}}{ 4\,\hslash^2 \; \mathcal{K}^2}\;\right) \,\right]
\end{equation}
\newline
where the BH oscillatory space-time parameters are expressed in terms of the BH surface  gravity 
 $\mathcal{K}$ Eq. (\ref{K}).
 $W_{sq}\left( \, X \,\mathcal{T} \,\right)$ clearly shows the circular structure of the quantum space- imaginary time instantons. The circular structure is manifest here without deformation because the $\alpha$ tail present for the coherent states is absent in this case.

\medskip

The coherent states for the quantum 
 gravitational instanton, here we explicitate for the BH, follow  similar expressions as Eqs.  (\ref{csdeSs}), (\ref{csdeSt}) but with the BH  factor $ 2\,\mathcal{K}$ : 
\begin{equation}\label{csinstBHs}
\Psi_{\alpha} \left( X\right)_{BH} \; = \; \, 
\left( \;\frac{i\,}{\pi \,}\;\right)^{1/4} \; \sqrt{\,2 \, \mathcal{K}\,}\;\, e^{-\frac{1}{2}\,\left\vert \,\alpha\,\right \vert
^{2}}\;\exp {\,\left[\;\sqrt{\,2}\, \;\alpha\; \mathcal{K}\,X \;(\, 1\; + \;i\,) 
 \; - \;2 \; i \;\mathcal{K}^2 \, X^{2}\;\right]}
\end{equation}
\begin{equation}\label{csinstBHt}
\Psi_{\alpha} \left( \,\mathcal{T}\,\right)_{BH} \; = \; \, \left(\;\frac{i\,}{\pi \,}\;\right)^{1/4} \;\sqrt{\frac{1}{\,2\,\hslash\;\mathcal{K}}}\;\, e^{-\frac{1}{2}\,\left\vert \,\alpha\,\right \vert
^{2}}\;
\exp {\,\left[\, - \frac{\alpha\,\mathcal{T}}{2\,\sqrt{\,2}\;\hslash\; \mathcal{K}} \;(\,1\; - \; i\,) \;  + \; \frac{ i\,\mathcal{T}^{2}}{8\,\hslash ^2\,\mathcal{K}^2}\;\right]}
\end{equation} 
\\
Therefore:
\begin{equation*} 
\left\vert \,\Psi_{\alpha} \left (\, X \,\right)_{BH}\,\right\vert^{2} \;\; + \; \;  \left\vert \,\Psi_{\alpha} \left (\,\mathcal{T}\,\right)_{BH}\,\right\vert^{2} \;\; = 
 \end{equation*}
 \begin{equation} \label{circleinst}
 = \;\, \frac{e^{-\,\left\vert
 \,\alpha\,\right\vert^{2}\,}}{\sqrt{\,\pi}}\,\left[\,2\;\mathcal{K}\,
 \exp{ \,[\;2\,c\,(\alpha)\,\mathcal{K}\,X\;]} \; + \; \frac{1}{ 2\;\hslash\; 
\mathcal{K}}\;
 \exp{\,[\;c (\alpha)\,\frac{\mathcal{T}}{ 2\;\hslash\; 
\mathcal{K}}\;]}\;\right] 
\end{equation}
$$
c\,(\alpha) \;\;=\;\;\sqrt{\,2}\;\,
(\, \text{Re}\,\alpha \;-\; \text{Im}\,\alpha\;)
$$
which reflects the quantum elliptic 
 (circular) structure of the space-imaginary time instanton. We define:
\be
\mathcal{R}_{\alpha} \,(\,X,\, \mathcal{T} \,)^2 \;\equiv \; \;\left\vert \,\Psi_{\alpha} \left (\, X  \,\right)_{BH}\,\right\vert^{2} \;\; + \; \;  \left\vert \,\Psi_{\alpha} \left (\,\mathcal{T}\,\right)_{BH}\,\right\vert^{2} 
\ee

We see that:
 For $\alpha \;=\;0$: 
 \be \label{alpha0bh}
\mathcal{R}_{0} \,(\,X,\, \mathcal{T} \,)^2 \;\; = \;\;\frac{1}{\sqrt{\,\pi}}\;\left[\;\,2\,\mathcal{K} \;+\; 
\frac{1}{2\,\hslash\,\mathcal{K}}\;\right]
\ee
For $(X, \mathcal{T})\, \rightarrow \,0 $ : 
\begin{equation}\label{XT0}
\mathcal{R}_{\alpha} \,(\,0 \,)^2 \;\;= \;
e^{-\,\left\vert
\,\alpha\,\right\vert^{2}\,}\; \mathcal{R}_{0}^{\,2}
\end{equation}

The origin is flurred or erased within a quantum circular core of radius $\mathcal{R}_{\alpha} \,(\,0\,)$. 
This confirms with a coherent state approach, the {\it regular} (non singular) quantum internal BH region obtained in Ref \cite{NSPRD2023} by using quantum Schwarschild-Kruskal coordinates. 

\medskip

At the Planck scale: $ (\,X, \,\mathcal{T}\,) \; \rightarrow (\,l_P,\,t_P\,) $, \; $\mathcal{K}\, \rightarrow \, \kappa_P = 1/(2\,l_P)$ : 

\be \label{planckbh}
\mathcal{R}_{\alpha} \,(\,l_P,\, t_P \,)^2 \;\; = \;\;
e^{\,[\, c\,(\,\alpha\,) \;-\; \left\vert
\,\alpha\,\right\vert^{2}  \;]} \, 
 \;\;\frac{1}{\sqrt{\,\pi}}\;\left[\;\frac{1}{l_P } \;+\; 
\frac{l_P}{\hslash}\;\right]
\ee
\\
As clearly seen, 
$\alpha = 0$ corresponds to the Planck scale (the onset scale in the trans-Planckian domaine). Consistently, the values $\alpha \neq 0$, \;
($ 0 <\,\alpha \, < \infty $),\, imply smaller sub-Planckian radii and more excited states, entering deeper in the quantum trans-Planckian region;  $\alpha$  can be very high but it is bounded,  as the quantum radius cannot be zero because of the quantum uncertainty, the notion of a {\it maximum} value $\alpha_{max}$ does appear related here to a minimal radius $\mathcal{R}_{min}$ due to the quantum uncertainty :
\begin{equation}\label{XT02}
\mathcal{R}_{min}\; \;= \;\;  \mathcal{R}_{\alpha-max} \,(\,0 \,) \;\;= \;\;
e^{\, - \left\vert
 \,\alpha_{max}\,\right\vert^{2} / 2} \;\;\mathcal{R}_{0} 
\end{equation}
 $\mathcal{R}_{0} \,(\,X,\, \mathcal{T} \,)$ Eq.(\ref{alpha0bh}) can be expressed in terms of the quantum uncertainties here :
 $$\Delta\,X \,= \, \frac{1}{\,\sqrt{\,2}\;2\;\mathcal{K}} \,,  \qquad \;\; \;\Delta\,\mathcal{T} \; = \;\sqrt{\,2}\;\hslash\;\mathcal{K}
 $$
\be
\mathcal{R}_{0} \,(\,X,\, T \,)^{\,2}\;\;=\;\;
\sqrt{\,\frac{2}{\pi}}\; \frac{1}{\hslash} \;\, \left[\; \Delta\,{T} \; + \; \Delta\,X \;\right]
\ee
\\
which is always non-zero because \,here \, $\Delta \,X\;\Delta \,T\;=\;\hslash \,/\,2$ , (minimal uncertainty). $ \alpha_{max}$ is thus given by:
\be
 \alpha_{max}^2 \;\; = \;\;2\;\log \;\left[\; \frac{1}{\pi}\;\left(\; \frac{1}{\Delta\,X} \; + \; \frac{2\,\Delta\,X} {\hslash}\;\right)\;\right]
\ee
\\
Consistently, at the Planck scale, we have :
$$(\Delta X)_P \; = \;  l_P\, /\,\sqrt{\,2} \, \qquad,\quad \, (\Delta T)_P \; = \; \hslash \,/\, (\,\sqrt{\,2}\;l_P\,)$$ 

\be \label{Rdelta}
\mathcal{R}_{0} \,(\,l_P,\, t_P \,)^{\,2}\;\;=\;\;
\frac{1}{\sqrt{\,\pi}}\;\left[\;\frac{1}{l_P } \;+\; 
\frac{l_P}{\hslash}\;\right]
\ee 
\\
which coincides with Eq. (\ref{planckbh}) for $\alpha \,=\, 0 $\,, as it mut be. 
It is remarkable how coherent states account for a consistent  quantum space-time description even at the Planckian and trans-Planckian scales.

Finally, the total {\bf discrete} $n$-states are given by the sum of the {\it even} and {\it odd} states. The BH $n$-states are similar to expressions Eqs.(\ref{stateH1}),(\ref{stateH1}) for deS space-time but, as we have seen,  the instanton corresponds to the normal (non-inverted) oscillator and then we have :  

\begin{equation}
\Psi \left(n\right)  \;\; = \;\; \Psi^{\left(+ \right)} \left(
2n\right) \;\; + \;\;\Psi^{ \left(- \right)} \left( 2n + 1\right) 
\end{equation}
where : 
\begin{equation} \label{stateBH1}
\Psi^{\left(+ \right)} \left(
2n\right)_{BH}\; = \;
\left\vert\,1\;+\;\mathcal{K}^4\,\right\vert^{1/4}\;\;\frac{\left(
\mathcal{K}^2 \right)^{2n}}{\sqrt{\,(2n)\,!}}
\end{equation}
\begin{equation} \label{stateBH2}
\Psi^{ \left(- \right)} \left( 2n + 1 \right)_{BH} \; = \;  \left\vert \,1 \;+\; \mathcal{K}^{4}\,\right\vert^{3/4} \;\, \frac{\left(\,\mathcal{K}^2 \,\right)^{2n + 1}} {\sqrt{\,\left(2n + 1\right)\,!} }
\end{equation}
\\
In addition, we take into account the quantum $\mathcal{K} $ discrete levels,  \cite{NSPRD2021},  \cite{NSPRD2023}: $\mathcal{K}_{Q\,n}\; = \;\sqrt{\,2 n}$\, \,({\it even} levels),\; and \; $\mathcal{K}_{Q\,n}\; = \;\sqrt{\,2n+1}$, ({\it odd} levels), which yields :

\begin{equation} \label{discreteBH1}
\Psi^{\left(+ \right)} \left(
2n \right)_{BH}\; = \;
\left \vert\; 1 \;+\; (2n)^2 \; \right \vert^{1/4}\;\;\frac{\,(\,2n \,)^{2n}}{ 2^{2n} \,\sqrt{\,(2n)\,!}}
\end{equation}
\begin{equation} \label{discreteBH2}
\Psi^{ \left(- \right)} \left( 2n + 1 \right)_{BH} \; = \;  [\;1 \; + \;(2n+1)^2\;]^{3/4} \;\,\frac{\left(\, 2n+1 \,\right)^{2n + 1}} { 2^{2n+1}\,\sqrt{\,\left(2n + 1\right)\,!} }
\end{equation}
\\
which completes all the states. The total covering is given by the sum of both $(\pm)$ states which cover the full Hilbert $Mp(2)$ space $\mathcal{H}_{1/4}\; \oplus\;\mathcal{H}_{3/4}$. This also shows that when considered in its full quantum discrete phase, quantum gravity must be a theory of pure numbers.

\section{Discussion}

It is interesting to discuss in this context the work by Ford Ref \cite{Ford1995} in which in a perturbative approach, quantum metric fluctuations can act as a regulator of the ultraviolet divergences of quantum fields. Metric fluctuations, as those due to gravitons in a quantum vacuum state, can modify the behavior of Green functions near the lightcone, smearing it  (and for instance in the one-loop electron self-energy). In other words,  gravitons in a quantum squeezed state could regulate ultraviolet divergences. 

\medskip

Our approach here is {\it non perturbative},  the light cone is fully quantum, the singularity is smeared out because of the non zero commutators $[\, X, T\; \text{or}\; P\,]$ or their quantum uncertainties, and a whole quantum region of finite curvature does appear : Thus, Ford perturbative proposal of smearing the light cone is fully realized in our approach non perturbatively within a whole quantum space-time description  (and quantum light cone). Our transverse spatial directions (or higher dimensions) to the lightcone are commutative  but can quantum fluctuate.

\medskip

Other points of comparison between our work here and  Ref \cite{Ford1995} are the following:

\medskip

(1) The line element in our case has in itself a quantum structure (group valued manifold) consequently giving rise to all the symmetries of both space-time and physically admissible states.
The shift metric in Ref \cite{Ford1995} is a perturbation that makes the resulting line element remain as the standard one plus the shift.

Compared to our line element in the super-Minkowski case (e.g. unperturbed metric), let us notice that the shift looks like the part of our line element that contains the complex $(\gamma)$ coefficients: the $B_1$ part of our $ds^2$ (Section III.A), for instance Eq. (\ref{A}), and Section (IV).

From the point of view of the obtained  solutions in both approaches: the same comparison is reflected in the Gaussian part of our solutions with the role of the complex parameters of the B1 part of our line element in the super-Minkowski case similar to the role of $\sigma_1$ in Eq.(10) of Ref \cite{Ford1995} .

\bigskip

(2) The shift in the interval (distance between events) of Ref \cite{Ford1995}  plays a similar role to the "fermionic" part $(B_1)$ of the line element described by us, not only from the point of view of the line element, but from the point of view of the Gaussian part of our solutions and the Green functions constructed by Ford in Ref \cite{Ford1995}: both the square norm of the metric coefficients of the fermionic part of our line element and the shift $\sigma_1$ in the case of Ref \cite{Ford1995}, locate the Gaussian function. 

Nevertheless, from a conceptual point of view, they are different because in our case the coefficients are not perturbations.

\bigskip

(3) In our approach the solution states are states with a certain spin content, in particular the case of spin 2 that correspond to the graviton field, and does not have singularity nor dynamic problems as we saw throughout the work.

\bigskip

(4) We could conclude that the Ford approach could be introduced or combined in our proposal but the action of the perturbation is screened by the metric coefficients of the fermionic part of the line element, given that these are not perturbations.
On the other hand, as a quantum result, the zitterbewegung does appear in our approach, which could approximately resemble to the fluctuations described in detail in Ford Ref \cite{Ford1995}.

\section{Remarks and Conclusions}

We have presented a non-perturbative group theory approach to describe quantum space-time and its states with new results both for quantum theory in its own and quantum gravity.

The results here provide further support to, 
and are  consistent with, the idea that a quantum theory of gravity must be
\textit{a finite theory}, (in the Wilson-Kadanoff sense),  which is more than a renormalizable theory, as discussed in Ref \cite{NSPRD2023}. And that a ultimate quantum theory of gravity must be a theory of {\it pure numbers}.

\medskip

{\bf(i)} We constructed here coherent and squeezed states of quantum space-time, in its continuum and discrete representations, and  for both, de Sitter and black hole space-times. They are naturally expressed in terms of the surface gravity.

\medskip

{\bf(ii)} We found that coherent and squeezed states of quantum space-time encompass the space-time behaviour in the 
semiclassical and classical de Sitter and  black hole regions,  
 they also exhibit high quantum phase space-time oscillations and they  account consistently  the Planckian and trans-Planckian scales.

\medskip

{\bf(iii)} We found the coherent states for the quantum space-imaginary time instantons,   black holes in particular, covering the  whole {\it complete} manifold including the  quantum central region, absent in the classical black hole instanton.

\medskip

{\bf(iv)} The Metaplectic group, the {\it complete} covering of SL(2C), plays an important role in providing the phase space symmetry of the coherent states and the complete Hilbert space whose two irreducible sectors span both the {\it odd} and {\it even} states, and whose total (symmetric or antisymmetric) covering guarantee  CPT symmetry and unitarity.

\medskip

The results presented here confirm that classical-quantum duality extended to gravity is a key part to understand the quantum gravity physical magnitudes, in particular the mass, and the space-time structure: classical-quantum gravity duality through the Planck scale (the crossing scale).  
 
\begin{itemize}
 \item{It is remarkable the power of coherent states in describing both continuum and discrete space-time, {\it even} in the Planckian and trans-Planckian scales:}

\item{ The continuum coherent state eigenvalue $\alpha = 0$, (and  the fundamental state $n = 0$ in the discrete representation), consistently correspond here to 
the Planck scale.  
Higher values of $\alpha $ consistently account for the smaller and sub-Planckian sizes and higher excitations in the quantum gravity domain.} 

\item{ We find a {\it maximum eigenvalue} $\alpha$ characterizing the coherent states due to the minimal non-zero quantum space radius  because of the  minimal quantum uncertainty $\Delta\,X \,\Delta\,T = \hslash\,/\,2$, in particular in the central and {\it regular}  black hole quantum region.} 

\item{In the quantum space-time description, there is no any space - time singularity as it must be. The consistent description  of such quantum scales by coherent states does appear as a result of the   space and time quantum uncertainties $(\, \Delta X \; \Delta T)$ and the classical-quantum gravity duality.}
\end{itemize}

A new  support to CPT and unitarity of the quantum gravity
theory does appear here through the Metaplectic group description.
In particular, recall that in semiclassical gravity, QFT in
curved space-times and its back reaction effects, the necessity of
considering  {\it complete} or CPT invariant states does appear in requiring unitarity of the theory , most investigated in the context of the  identification of space-time ("IST"), \cite{SanchezCargese1987}, \cite{NGSWhitingNPB1987}, \cite%
{NGSDomechLevinasIJMPA1987}, \cite{tHooft1}, \cite{tHooft2022},
\cite{StraussWhitingFranzenCQG2020}. See also \cite{kumar} for a recent account without IST.
We
have not used any identification of space-time ("IST") here, but our results here, as those in  \cite{NSPRD2021}, \cite{NSPRD2023}, \cite{Sanchez2019}, support CPT and IST in the full quantum theory. 

In semiclassical gravity, the symmetric (or antisymmetric) QFT provides a CPT symmetry of the theory. In the euclidean (imaginary time) manifold, the
different causally disconnected regions 
became automatically identified. The quantum instanton contains in addition the central regular constant curvature region of Planck scale radius not covered by  the classical instanton. The coherent state instanton remarkably accounts for this quantum gravity feature and determines the radius being \;$$\mathcal{R}_{0} \,(\,l_P,\, t_P \,)^{\,2}\;\;=\;\;
\frac{1}{\sqrt{\,\pi}}\;\left[\;\frac{1}{l_P } \;+\; 
\frac{l_P}{\hslash}\;\right], $$\; 
$l_P$ being the Planck length. The origin is flurried or smoothed within this constant  and bounded curvature region.
 
\bigskip

The results of this paper confirm that the quantum de Sitter vacuum and the quantum interior region of
the black hole are  both of the same nature:  totally regular \textbf{without any
curvature singularity} and of constant curvature. These results provide too a quantum  space-time support to the effective or phenomenological models, \cite{frolov1}, \cite{brustein1}, \cite{brustein2}, describing the BH interior  
as a de Sitter core of bounded curvature and being totally regular. 

\bigskip 

The results of this paper are expected to provide new insights to explore the quantum space-time structure and its signals,
being from black holes, the gravitational wave domaine and the high energy domain, or the de Sitter primordial phases (inflation and before inflation),  cosmological structures and  the late de Sitter cosmological vacuum  (today dark energy), Refs \cite{LISA} to \cite{LSST}. The classical-quantum gravity duality allows that signals in the quantum gravity (trans-Planckian) domaine  do appear as low energy effects in the semiclassical/classical universe today. 

\medskip

Notice too that the quantum gravity regions, the black hole interiors for instance, are present in  all black holes of {\it all} masses, including the most macroscopic and astrophysical black holes.

\medskip

Interestingly, the results of this paper can also provide with the coherent states of quantum space-time a {\it quantum optics} of the space-time and its tests, or find analogous of it, in the wave packet type and laboratory experiments.

\newpage

\textbf{ACKNOWLEDGEMENTS}

\bigskip

\bigskip

DJCL acknowledges institutionally to the CONICET and  the Keldysh Institute of
Applied Mathematics; and to the support of the Moscow Center of Fundamental
and Applied Mathematics, Agreement with the Ministry of Science and Higher
Education of the Russian Federation, No. 075-15-2022-283. NGS thanks useful discussions with Gerard 't Hooft, Roger Penrose and Adam Riess on various occasions.

\end{document}